\documentclass[11pt,a4paper]{article}

\usepackage[no-natbib-sort]{my-jheppub}

\usepackage{feynmp}
\usepackage[nodayofweek]{date time}

\def \fo {{\textstyle { 1 \ov 4}}}
\def \la {\label}
\newcommand \foot [1] {\footnote{#1\vspace{2pt}}}
\newcommand \rf [1] {(\ref{#1})}

\def \ha {{ \textstyle{1 \over 2}}}

\def \ci{\cite}
\def \del{ \partial}
\def\ov{\over}
\def\no{\nonumber} 

\def \p {\phi}
 
\def \te {\textstyle} 
\def \ed  {\end{document} }
\def \fo {{ \textstyle{1 \over 4}}}

\begin{document}

\date{\currenttime}


\title{On quantum corrections\\ in  higher-spin theory   in flat space } 

\author{Dmitry Ponomarev\footnote{d.ponomarev@imperial.ac.uk}}
\author{\ and \  Arkady  A. Tseytlin\footnote{Also at Lebedev Institute, Moscow. \ \ tseytlin@imperial.ac.uk}}

\affiliation{Theoretical physics group, Blackett Laboratory, Imperial College London,   SW7 2AZ, U.K.}


\abstract{
We consider  an  interacting theory of  an infinite tower  of  massless  higher-spin fields in flat space with  cubic vertices   
and their coupling constants  found  previously by Metsaev.  We compute the one-loop 
 bubble diagram part of  the self-energy of the  spin 0  member of the tower    by summing up  all 
 higher-spin loop contributions.  We find that the result contains an  exponentially  UV divergent  part   and we discuss  how it  could  be cancelled   by a tadpole contribution  depending on  yet to be determined  quartic interaction vertex. 
We also compute the tree-level four-scalar scattering amplitude  due to all higher-spin exchanges  and 
discuss its inconsistency with the BCFW constructibility condition. We   comment  on possible  relation to similar computations in AdS   background  in connection  with    AdS/CFT.
}

\unitlength = 1mm

\today
\date {}
\begin{flushright}\small{Imperial-TP-DP-2016-{01}}\end{flushright}

\maketitle

\def \lR {L}

\def \ll {{\cal \ell}}

\def \be {\begin{equation}}\def \ee {\end{equation}}

\def \ads {AdS$_4$\ }
\def \iffa {\iffalse} 

\def \ed {\bibliography{loopsbib}{}
\bibliographystyle{utphys} \end{document}}

\section{Introduction}

A possibility of   an interacting theory of an infinite tower of 
massless higher spins in flat space is an old question with various no-go theorems 
prohibiting the existence of minimal  (low-derivative) couplings  or
  long-distance  interactions  (see  \ci{Bekaert:2010hw}
 for a review).
 At the same time,  non-trivial cubic  vertices    containing higher derivatives 
  were  constructed in the   past   using   various 
  approaches  \cite{Bengtsson:1983pd,Berends:1984wp,Berends:1984rq,
 Bengtsson:1986kh,
 Metsaev:1991mt,Metsaev:1991nb,Metsaev:1993ap,Metsaev:2005ar,
 Fotopoulos:2008ka,Zinoviev:2008ck,Boulanger:2008tg,Manvelyan:2010jr,Sagnotti:2010at,Fotopoulos:2010ay,
 Manvelyan:2010je}.\foot{
 How some constraints of  no-go theorems  may be  avoided is discussed  in  \ci{Bekaert:2010hw,Taronna:2011kt,Sagnotti:2013bha}.
 The Coleman-Mandula theorem that prohibits existence of higher spin conserved charges 
 assumes finite  number of particles of mass  below certain scale and analyticity of the amplitudes. 
 Weinberg's  soft theorem  \ci{Weinberg:1995mt}  is a requirement  of linearized gauge invariance  with respect to the 
 spin $s$ leg in the amplitude.
 In general,  the soft theorem  constrains  
  $s$-$s'$-$s'$   couplings  with
 minimal number  $s$ of derivatives in the vertex. It  
   imposes   conditions  on  scattering amplitudes 
 with at least one higher spin $s$  particle    on an external line  which 
 may be  equivalent to  conservation of 
 higher spin charges (which are higher than quadratic in external momentum and thus allow only 
  elastic scattering with permutations of momenta). 
  It  leads to non-trivial constraints only   if exchanged spins are 
  lower than external ones  \ci{Taronna:2011kt}. 
 Constraints on cubic couplings  based on the assumption   of  BCFW 
 constructibility \ci{Britto:2005fq}  
 applied to   massless  4-point scattering amplitude 
    \ci{Benincasa:2007xk,Benincasa:2011pg,McGady:2013sga}
  may not apply if one 
  allows   for some non-locality 
  of the 4-point  vertex  in a theory  that  contains  infinite sum over  
  spins/derivatives. It is  likely  that  the condition of BCFW constructibility is stronger than that of  the existence of an interacting theory   constructed  from a free theory via gauge symmetry  deformation or Noether method (see also discussion below). 
 }
 
 The allowed cubic vertices $\del^n \p_{s_1} \p_{s_2} \p_{s_3} $   have  the number of derivatives  $n$ 
 constrained by $  s_2+s_3 -s_1 \leq  n \leq s_2+s_3 +s_1$ (assuming $ s_1 \leq s_2\leq s_3$).
 Remarkably, the coupling constants in these vertices  can be fixed  in terms of a   single  dimensional 
  constant $\ll$ \ci{Metsaev:1991mt}.
 For example, for the highest-derivative 
  vertex one gets   $ g_{s_1s_2s_3} \sim  { \ll^{ s_1+s_2+s_3-1} \ov (s_1+s_2+s_3-1)!} $.
 Assuming that a Noether  deformation  procedure  can  completely determine   also the quartic and higher vertices, 
 one may conjecture the  existence of a theory containing an   infinite tower of  massless Fronsdal  fields $\p_s$  ($s=0,1,2,...,\infty$)
 with   an  action depending on one dimensionless  coupling $g$ and  one 
 dimensional parameter $\ll$  and having the following  structure\foot{Here we will consider   the case of 4d space-time (so that 
 $\p_s$ are assumed to have  dimension  length$^{-1}$)  but most of the discussion below   will  be  true also in any dimension.
 In $d=4$  the light-cone gauge vertices of  \ci{Metsaev:1991mt} 
   that admit a local covariant  generalization  contain only  two structures:   with  maximal 
    $s_2+s_3 +s_1$   and with  minimal $s_2+s_3 -s_1$  numbers of derivatives.
    } 
 \begin{align} \nonumber 
 S_{\rm flat} &= {1\ov g^2} \int d^4 x  \Big[ \sum_s  \p_s \del^2 \p_s  +   \sum  \ll^{n-1}  \del^n \p_{s_1} \p_{s_2} \p_{s_3}
 + \sum \ll^{k-2}  \del^k \p^4 + ...\Big] \\
 &\to\   \int d^4 x  \Big[ \sum_s  \p_s \del^2 \p_s  +  g \sum  \ll^{n-1}  \del^n \p_{s_1} \p_{s_2} \p_{s_3}
 + g^2 \sum \ll^{k-2}  \del^k \p^4 + ...\Big] \ . 
 \la{1}
  \end{align} 
Here $g$ controls the expansion in number of  fields (and also loop expansion) while  $\ll$ sets up an effective scale (i.e. appears together 
with derivatives or momenta). Such a  theory    is  effectively non-local   --   the number of derivatives in the interaction terms  is unbounded  
 as spins can take any  value up to infinity.

One may wonder   why such  an  unusual   theory (assuming it indeed  exists)  may be of any  interest. 
One reason is that it may  have hidden simplicity  due to its expected   large  gauge (and global) higher spin 
 symmetry.  
  For example, the free 
theory $\sum^\infty_{s=0} \int d^4x \ \p_s \del^2 \p_s$   turns out to have  zero total number of  dynamical degrees of freedom  and thus  trivial partition function $Z=1$ \ci{Beccaria:2015vaa}.\foot{In contrast to supersymmetric  
theories  here $Z=1$   even at  finite temperature (all bosonic fields have the same statistics) 
 so   a closer analogy is with a topological field theory. 
 A non-trivial generalization of $Z$ to quotients of flat space in 
  the presence of angular potentials was discussed in \ci{Campoleoni:2015qrh}.
 } 
This is true  if one uses a particular prescription of summation over spins that should   be consistent with underlying symmetry of the theory. 
One may then expect  that   under such summation  prescription other quantum corrections may also be simple.\foot{Similar simplifications 
were observed in conformal higher-spin theory  \ci{Tseytlin:2013jya,Beccaria:2015vaa,Joung:2015eny}.
The existence of higher spin conserved   charges  may also have  drastic 
 consequences   for the S-matrix with higher spin particles on external lines (as  in the case of integrable theories in 2 dimensions \ci{Parke:1980ki}).}
For example, 
despite having a dimensional coupling $\ll$  the  theory \rf{1}  may  actually be UV finite. This would  be analogous to what happens in string theory   viewed  as a collection  of a few massless  and an infinite set of massive higher spin fields  where  a particular prescription of summation over all  contributions implied by the  underlying world-sheet   formulation   leads to the UV finiteness of  scattering amplitudes.\foot{At the same time,  it   seems  unlikely  that  \rf{1}  may  be consistently related to  a zero-tension limit  of  the 
bosonic string theory   combined with a truncation to the  leading Regge  trajectory:   zero-tension limit does  not
appear to be  well-defined in flat space and, moreover, the tower of fields on the  leading Regge trajectory may 
not lead to a  UV-consistent    theory on its own.
 }

Another motivation to study \rf{1} is  its  possible relation 
to the  massless   higher-spin theory in  AdS  space. The existence of the consistent  cubic  couplings (containing  also  low-derivative  or minimal-coupling 
parts and  avoiding the no-go theorems) was first pointed out in \ci{Fradkin:1986qy,Fradkin:1987ks}. 
One may  conjecture  that  an elimination of an infinite set 
 of the auxiliary fields  present  in  the  non-linear  Vasiliev's equations \ci{Vasiliev:1990en,Vasiliev:2003ev} 
 expanded   near the AdS vacuum  may 
lead to an action for the  tower  of  physical  massless  Fronsdal   fields   that has a structure  similar  to that of \rf{1} 
(cf. \ci{Giombi:2009wh,Giombi:2012ms,Boulanger:2015ova})
 \be\la{2} 
 S_{\rm AdS}= {1\ov g^2}\int d^4 x  \Big[ \sum_s  \p_s ( \nabla^2+...) \p_s  +   \sum  \lR^{n-1}  \nabla^n \p_{s_1} \p_{s_2} \p_{s_3}
 + \sum \lR^{k-2}  \nabla^k \p^4 + ...\Big] \ . \ee 
Here $\lR$  is the AdS  radius   and  the cubic vertices  now  contain also the low-derivative (``minimal-coupling") 
tail  of terms (e.g.  $n=1,2,...$). 
To understand  a possible   reason  for their  presence let  us  imagine  that  the action \rf{1}    admits
 a generalization to  curved 
background  where the  flat-space metric   is   replaced by a curved one $g_{ab}$   and  the flat   derivatives $\del_a$ by the 
covariant ones $\nabla_a$. 
Then in addition to the cubic terms  $\ll^{n-1} \nabla^n \p_{s_1} \p_{s_2} \p_{s_3}$ 
as in \rf{1} with  $n \geq s_2+s_3-s_1$  one may    also have the terms 
of the same dimension   with  less   derivatives  but  with extra powers of the curvature, 
$\ll^{n-1} R^{k} \nabla^{n-2k}  \p_{s_1} \p_{s_2} \p_{s_3}$, i.e.  $\ll^{n-1} L^{-2k}  \nabla^{n-2k}  \p_{s_1} \p_{s_2} \p_{s_3}$ 
   in the case of the  AdS space with $R\sim L^{-2}$.
    In the action \rf{2} corresponding to the Vasiliev-type theory there is  just 
     one dimensional scale, i.e. $\ll$ is effectively  identified with  the curvature scale 
     $L$ and thus    the flat space limit ($\lR\to \infty$)   is  formally singular.\foot{Here we assumed that the curved 
    metric  (e.g., $ds^2_{\rm AdS} = du^2 + e^{-u/L} dx^m dx_m$)  has a smooth flat  space 
    limit without  need  to rescale the coordinates. In general, one may contemplate taking a flat  space  limit that involves some singular   rescalings of the fields, coordinates and coupling  constants 
    (as, e.g.,  in \ci{Boulanger:2008tg,Joung:2011ww}).
     It is not clear   if such a limit may actually exist   beyond the cubic 
 interaction level   given that quartic vertices  may  contain  sums over 
  all orders in derivatives for fixed external spins.
    }

 While a naive    flat-space 
limit  that  leads  from \rf{2} to \rf{1}   may not exist, there may be still some formal procedure of  relating 
the two actions    in which 
these  low-derivative couplings  would   decouple    with    $\lR$ in \rf{2}  being effectively 
 replaced by another scale $\ll$ in \rf{1}. 
   One   might think  that  as  the leading  short-distance behaviour    should   be controlled by the highest  derivative terms,  
  one may   then  expect  that the UV properties 
of a  flat-space  and a   curved-space (e.g.  AdS)  theories     may  be similar.  
This, however,   may not   be  true   in the present  context of an effectively non-local theory 
(containing all powers of derivatives in the vertices due to the 
summation over an infinite set of higher spin fields).
Still,   the study of a  simpler theory \rf{1}   may 
shed  some  light on some  properties of \rf{2}.

The definition of  a  quantum  theory   with  an infinite set of   fundamental fields   like the one in   \rf{1} or \rf{2}  
is  a priori ambiguous  as the  sum of the individual field  contributions over all spins  may be divergent. 
This ambiguity is to be fixed in  a way that is consistent  with preservation of underlying higher spin symmetry as 
was recently discussed in  \cite{Giombi:2013yva,Giombi:2013fka,Tseytlin:2013jya,Giombi:2014iua,Giombi:2014yra,
Beccaria:2014jxa,Beccaria:2014xda,Joung:2015eny}.
  In particular,  under a special  summation  prescription,  the    free-field 
partition function  of the  AdS theory  \rf{2}   is trivial, i.e.   $Z=1$ \ci{Giombi:2013fka,Giombi:2014iua},   just  
 as in the  flat space theory  \rf{1} \ci{Beccaria:2015vaa}.

 This definition   should   be  consistent with the  vectorial AdS/CFT   duality
  \cite{
  Klebanov:2002ja,Leigh:2003gk,Sezgin:2003pt}  
   between the massless   higher spin 
 theory in AdS$_{d+1}$   and  the  singlet sector of  the free $U(N)$  or $O(N)$   scalar CFT in $\mathbb R^d$
 which is also controlled by the underlying higher spin symmetry. 
 This duality   
 was tested at the tree level for some  3-point functions 
\cite{Giombi:2009wh,Giombi:2010vg} which are essentially fixed by the unbroken higher spin 
symmetry \ci{Maldacena:2011jn}.\foot{Also, 
 all $n$-point functions of the free CFT have been identified
with suitable invariants in the Vasiliev theory in \cite{Colombo:2012jx,Didenko:2012tv,Gelfond:2013xt}.}
Let us note that some  cubic vertices for the  Fronsdal   fields derived from the Vasiliev's theory 
upon   elimination of auxiliary fields appear to be   formally divergent 
 \cite{Giombi:2009wh,Giombi:2010vg,Boulanger:2015ova} and thus   require
 special definition 
 even before summation over all physical spins
  (cf.  \cite{Vasiliev:2016xui}).

  Alternatively,  one can try to reconstruct   the cubic   and quartic 
  action \rf{2}   (with $g^2= 1/N$)  by requiring that it  should 
   reproduce   the   boundary  CFT correlators    at the tree level  
 \cite{Bekaert:2014cea,Bekaert:2015tva,Sleight:2016dba} (for  related earlier work  see
\cite{Petkou:2003zz}).
Then    a  crucial  test of the duality  will be  to check    that all   quantum corrections to
 the  ``on-shell"  value of  the 
effective  action of  the theory \rf{2}    should vanish  since the   correlators  of conserved currents 
in  the free  boundary CFT  do not receive $1/N$ corrections, i.e. are given exactly by their large $N$ values.\foot{A priori it is possible 
that quantum corrections to the effective action of the theory \rf{2} 
could  be non-vanishing but having   such  special ``local" form  that they do not contribute to 
derivatives over the  boundary sources  taken  at separated  points. 
At the same time,  this seems unlikely as  quantum corrections should be controlled 
 by the higher spin symmetry that   should constrain  also possible contact terms.}  
Assuming that  the matching of correlators  at separated points  may  be extended also to integrated correlators,  
this    suggests, in particular,  that the  vacuum  partition function of the AdS theory \rf{2}   should   vanish not just 
at the leading one-loop order \ci{Giombi:2013fka}   but also  to   all orders  in  $g= {1/ \sqrt N}$  expansion. 

Another  non-trivial  quantum test   would   be  the  demonstration of the 
vanishing of the 1-loop  correction to the  spin $s$  field 2-point  function  as 
 the  2-point  functions or 
 dimensions  of the currents in the 
boundary CFT should  not be $1/N$-corrected in the  case of unbroken higher spin symmetry. 
This    ``self-energy"    correction  is given by the sum of  the two types of the one-loop Witten  diagrams:  the    bubble 
 diagram (with two bulk-to-bulk propagators and two cubic vertices) 
and    the  tadpole  diagram  determined by the quartic vertex in \rf{2}.\foot{Some one-loop corrections 
  to propagators in AdS were computed previously  for spin 2 in \cite{Porrati:2001db,Duff:2004wh}
   and for higher spins in   \cite{Manvelyan:2008ks}.  
 One-loop computations  based on Mellin representation  were performed in \cite{Penedones:2010ue}.
  Related discussions from higher spin AdS/CFT
    perspective appeared  in  \cite{Skvortsov:2015pea,Giombi:2016hkj,Hikida:2016wqj,Bashmakov:2016pcg}.
   Assuming this cancellation,  the  AdS/CFT matching in the case of the broken  higher spin symmetry 
   was discussed in \ci{Giombi:2011ya}.
   }

With this motivation  in mind here we  will address a  simpler question about  the one-loop  bubble-diagram   part  of 
the self-energy  correction  in the  higher spin  flat space theory \rf{1}.
Certain features of the flat-space 
result should  be similar to the ones  in the AdS  case, at least in what concerns the leading UV behaviour. 
Our aim will be to extract  the   UV  divergence   of the  bubble  diagram 
 and to see if  it can be    cancelled against a tadpole contribution  coming from  (yet unknown)  4-point  vertex.    
 
 For simplicity,  we shall consider  the case when only  the 
 scalar  particle (spin 0   member of the higher spin tower)   appears on the two external lines of the self-energy diagram. 
 In this case the  bubble graph contribution 
  is determined by the  3-point  vertex   0-$s_2$-$s_3$   (containing  $n=s_2 + s_3$  derivatives) 
   which is essentially  unique. 
      We shall  use   the  explicit   value of its coefficient 
 $\sim { \ll^{s_2  + s_3 -1} \ov { (s_2 + s_3 -1)!}}$   found in  \ci{Metsaev:1991mt}   and  perform the summation over all   spins
   in the loop.\foot{This cubic  vertex  can be found, e.g.,  by requiring   the gauge invariance
of the full  non-linear action to the lowest order in the coupling  $g$. At this order
the  coupling constants of individual vertices for different spins
  are independent.  By requiring
gauge invariance to the next $g^2$ order, these coupling constants can all be expressed in terms
of a unique dimensionful parameter  $\ell$  as in \rf{1}.
An equivalent analysis  was carried out in  flat 4d space in 
 the light-cone approach in   \cite{Metsaev:1991mt}.
In AdS similar relations were found within  the  Fradkin-Vasiliev approach \cite{Fradkin:1987ks,Vasilev:2011xf,Boulanger:2012dx}
by imposing the Jacobi identity for the gauge 
symmetry deformations  \cite{Fradkin:1986ka,Boulanger:2013zza}.}  
   The one-loop diagram will then  be   quadratic  in dimensionless coupling $g$  in \rf{1}   and  given by 
   the virtual momentum   integral   with non-trivial dependence on  the product of $\ll$ with external momentum. 
  
  To find the full result for the $g^2$ correction to self-energy diagram one is to add  also   the ghost-loop  contribution 
  and   the tadpole graph contribution. The latter requires the knowledge of the  0-0-$s$-$s$ 
 quartic   vertex   (and similar one for the ghosts).
 The problem of determining quartic interactions of massless higher spins in flat  and AdS spaces 
  was addressed using  different approaches in, e.g., 
    \cite{Vasiliev:1989yr,Metsaev:1991mt,Sagnotti:2010at,Polyakov:2010sk,Taronna:2011kt,Dempster:2012vw,Buchbinder:2015apa,
    Bekaert:2015tva}, 
    but its full  conclusive solution  is yet to be found. While being unable to determine  
     the tadpole contribution explicitly, 
       here we shall still comment on its expected UV   behaviour required to cancel the UV divergent part of the bubble graph. We shall also note  that the tadpole   contribution  is not expected to  alter the non-trivial (non-analytic) 
    external  momentum dependence  coming from  the bubble graph.

The rest of this   paper is organized as follows. 
In section 2  we shall  define  the  free  Fronsdal action for the  totally symmetric massless  higher spin fields in flat  $d$ dimensions
and impose the de Donder gauge.  
In section 3  we shall discuss the structure of  the 0-$s_2$-$s_3$ cubic vertices  required for subsequent computations. We shall
  also consider the leading term in the  deformation of the  gauge transformations due to the presence of the cubic interactions 
 and thus  determine the   corresponding  quadratic and cubic  terms in the ghost action. 
In section 4 we shall review the derivation of the  higher spin propagator in the de Donder gauge and 
describe the resulting Feynman rules. 

Section 5 will be devoted to the computation of the tree-level 4-scalar scattering  amplitude. We shall first explicitly compute the  exchange  part of the  amplitude and then comment on possible contribution of the 4-scalar  ``contact"  vertex. 
 In  section  6  we shall compute the  bubble diagram   contribution to the one-loop 
 scalar self-energy corrections and   discuss its   UV behaviour.  We shall then
  discuss a possibility that the UV divergence of the bubble  diagram may be cancelled  against 
   a  tadpole graph contribution determined  by the 4-point vertex. 
   Some concluding remarks will be made in section 7. 
   
In Appendix  A  we shall explain the relation   between  the covariant  cubic  vertex we use  and the  light-cone gauge cubic  vertex
 in  \cite{Metsaev:1991mt}. In Appendix B we shall  discuss 
  whether these vertices  are consistent
 with the BCFW constructibility condition (see also  comments at the end of section 5).
 Appendix C will  contain some details of the computation of sums over spins  in section 6. 


\section{Free higher spin action}

To 
make expressions more compact we shall  represent the totally symmetric 
higher spin tensor  fields   by 
\begin{equation}\la{0}
\phi_s(x,u) =\te \frac{1}{s!} \phi_s^{a_1\dots a_s}(x)\, u_{a_1}  \dots u_{a_s}\ , 
\end{equation}
where $u_a$ is an arbitrary constant vector. 
Then the  Fronsdal action \cite{Fronsdal:1978rb}  may be written as\foot{The action (\ref{1b}) is canonically normalised  and its  overall
sign is chosen appropriately to ensure positive energy of field fluctuations
for the mostly plus Minkowski metric $\eta=\rm{diag}(-,+,+,\dots,+)$.} 
\begin{equation}
\label{1b}
S^{(2)}[\phi_s]=\te \frac{s!}{2}\int d^dx \left[\phi_s(x,\partial_u) \,\hat T \hat{\cal F} \;\phi_s(x,u)\right]_{u=0},
\end{equation}
where 
\begin{equation}
\label{2b}
\hat T \equiv 1-\fo u^2\partial_u^2,\qquad
\hat{\cal F} \equiv \partial_x^2 - (u\cdot \partial_x) \,\hat D, \qquad \hat D\equiv 
(\partial_x \cdot \partial_u)-\ha
(u\cdot \partial_x) \partial^2_u  \ , 
\end{equation}
and the off-shell  field $\phi_s$ is  assumed to be double-traceless, i.e.   satisfying  ($\partial_u^2\equiv  \eta^{ab}{\del\ov \del u^a} {\del\ov \del u^b}$)
\begin{equation}\la{01}
(\partial_u^2)^2 \phi_s (x,u)=0\ .
\end{equation}
As in the second line of  \rf{1} we assume that the dimensionless  coupling $g$ is absorbed into  $\p_s$   so 
that it will appear in the interaction vertices. 

The equation of motion for (\ref{1b}) is 
\begin{equation}
\label{eom01}
\frac{\delta S^{(2)}[\phi_s]}{\delta \phi_s} = \hat T \hat{\cal F}\;\phi_s(x,u)\approx 0.
\end{equation}
Here and in what follows 
we use the symbol $\approx$ to denote the    equalities that hold modulo terms proportional to the free equations
of motion. 
By noting that $\hat T$ is invertible
\begin{equation}
\label{12proc2}
\hat T^{-1}\equiv  1- \frac{1}{2d+4u\cdot \partial_u -12}u^2 \partial_{u}^2\ , \qquad 
\hat T^{-1} \hat T\,
\phi_{s}(x,u) = \hat T \,\hat T^{-1}\phi_{s}(x,u) 
= \phi_{s}(x,u)\ ,
\end{equation}
one finds that the free equations \rf{eom01}
 can be equivalently rewritten as
\begin{equation}
\label{eom}
\hat{\cal F} \;\phi(x,u)\approx 0.
\end{equation}
Note, that in (\ref{12proc2}) we kept only those terms in $\hat T^{-1}$ that do not annihilate
double-traceless tensors, i.e. do not contain $(\partial_u^2)^{n}$ with $n>1$   (cf. \rf{01}).

The Fronsdal action \rf{1b}  is invariant under the  gauge transformations 
\begin{equation}
\label{3}
\delta^{(0)}_{s} \phi_s(x,u) = (u\cdot \partial_x) \varepsilon_{s-1}(x,u),
\end{equation}
with  the traceless  gauge parameter  $\varepsilon_{s-1}(x,u)$ 
\begin{equation} \la{03}
 \varepsilon_{s-1}(x,u) \equiv  \frac{1}{(s-1)!} \varepsilon_{s-1}^{a_1\dots a_{s-1}}(x)\, u_{a_1} \dots u_{a_{s-1}}\ , \qquad \qquad 
 \partial^2_u\;\varepsilon_{s-1}(x,u)=0
\ . 
\end{equation}
It is convenient to impose the de Donder gauge 
\begin{equation}
\label{4}
\hat D\,\phi_s(x,u)=0 \ . 
\end{equation}
In this gauge the Fronsdal operator  $\hat{\cal F}$ in  (\ref{2b})  simplifies so that  the 
action \rf{1b} becomes 
\begin{equation}
\label{5}
S^{(2)}[\phi_s]=\te \frac{s!}{2}\int  d^d x  \left[ \phi_s(x,\partial_u)\,\hat T\,\partial_x^2\, 
  \phi_s(x,u)\right]_{u=0}\ , 
\end{equation}
while the equations of motion \rf{eom}    take the form 
\begin{equation}
\label{7}
\Box   \phi_s(x,u)\approx 0\ , \qquad \qquad \Box=  \partial_x^2 \ . 
\end{equation}

\section{Cubic interaction vertices} 

In this section we shall present  cubic vertices for  the physical fields $\p_s$ 
and the ghosts corresponding to the de Donder gauge \rf{4}. 
 To construct  cubic 
vertices in the covariant form one usually starts by specifying their traceless transverse 
parts. 
Then these vertices can be completed to full-fledged off-shell ones
  \cite{Manvelyan:2010jr,Sagnotti:2010at,Fotopoulos:2010ay,Manvelyan:2010je}.
 For our purposes it is not necessary to face  the difficulties inherent to general off-shell interactions as it 
 suffices to know the vertices in the  de Donder gauge. 
Moreover,  as  here we will be interested    in  computing   diagrams with  only spin 0 particles on 
 external lines    we may restrict consideration to  cubic   vertices with one of the  fields having  $s=0$. 
In this  case   it turns out  that  the traceless-transverse vertices
  give already  the  consistent  vertices in the de Donder gauge, i.e.  do not require any completion.\foot{
 The observation  that the traceless-transverse
vertices  are sufficient to find the  consistent vertices in the de Donder gauge 
 was already made  in  \cite{Manvelyan:2010jr,Sagnotti:2010at,Fotopoulos:2010ay,Manvelyan:2010je}.
 In the most general case in the de Donder 
gauge apart from traceless-transverse  contribution there is also a term  which is 
cubic in traces   \cite{Manvelyan:2010jr}. In our case it is absent and  this  is indeed an extra
 simplification that we use.
 }

\subsection{Deformation  of the free action}

Adding cubic interaction part $S^{(3)} $   to the free action $S^{(2)} $    and requiring gauge 
 invariance of the combined  non-linear action  gives at the first non-trivial order  the  condition 
\begin{equation}
\label{11}
\delta^{(0)} S^{(3)} + \delta^{(1)} S^{(2)} =0\ ,
\end{equation}
where 
 $\delta^{(1)}$ is a 
deformation of the gauge
transformation (\ref{3})  which is linear in  the fields $\p_s$. 
The first term in (\ref{11}) thus 
vanishes modulo the free equations of motion \rf{eom01}, i.e. 
\begin{equation}
\label{11imp}
\delta^{(0)} S^{(3)} \approx 0\ .
\end{equation}
Once the cubic vertex is found, the associated deformation of the gauge transformation can be 
extracted from (\ref{11}).

The traceless-transverse part of the cubic vertex involving the spin 0  field and  any  two other 
higher spin fields can be written as \cite{Manvelyan:2010jr,Sagnotti:2010at,Manvelyan:2010je,Joung:2011ww}
\begin{align}
\notag
S^{(3)}[\phi_0,\phi_{s_2},\phi_{s_3}] = g_{0s_2s_3}\int d^dx \Big[(\partial_{u_2} \cdot \partial_{x_{31}})^{s_2}
&(\partial_{u_3} \cdot \partial_{x_{12}})^{s_3}\\
\label{8}
& \times \phi_0(x_1)\phi_{s_2}(x_2,u_2)\phi_{s_3}(x_3,u_3)\Big]_{\substack{{u_i=0}\\{x_i=x}}}\ ,
\end{align}
where
$ \partial_{x_{ij}} \equiv \partial_{x_i}-\partial_{x_j} $
and $g_{0s_2s_3}$ is a 
coupling constant.
To show that it gives a consistent vertex in the de Donder gauge,
let us  verify that it satisfies (\ref{11imp}). Using that 
\begin{equation}
[\partial_{u_2}\cdot \partial_{x_{31}}, u_2 \cdot \partial_{x_2}  ] = (\partial_{x_3} -\partial_{x_{1}})\cdot \partial_{x_2} = -\partial^2_{x_3} + \partial^2_{x_1} +\text{t.d.}
\end{equation}
where ``t.d."  stands for a  total derivative term, we find 
\begin{align}
\notag
\delta_{{s_2}}^{(0)} \phi_{s_2}\frac{\delta S^{(3)}[\phi_{0},\phi_{s_2},\phi_{s_3}]}{\delta \phi_{s_2}}
 =& g_{0s_2s_3}\,   s_2\int d^dx \Big[ (\partial_{u_2} \cdot \partial_{x_{31}})^{s_2-1}
(\partial_{u_3} \cdot \partial_{x_{12}})^{s_3}\, (\partial^2_{x_1}-\partial^2_{x_3})\\
\label{9}
&\qquad\qquad   \times \phi_0(x_1)\,\varepsilon_{s_2-1}(x_2,u_2)\,\phi_{s_3}(x_3,u_3)\Big]_{\substack{{u_i=0}\\{x_i=x}}}\approx 0\ .
\end{align}
This  leading-order deformation analysis  fixes the structure of the 
cubic vertices but leaves the coupling constants 
$g_{0s_2s_3}$ in \rf{8} undetermined. 
This happens because the gauge invariance conditions
 (\ref{11}), (\ref{11imp}) are linear in  the deformation $S^{(3)}$.
To find   $g_{0s_2s_3}$ one needs to  consider  higher-order deformations  and 
to solve analogous  higher-order  constraints  
 which  
become non-linear in the fields. At the  next order  one gets the condition 
\begin{equation}
\label{nquartic}
\delta^{(0)} S^{(4)} + \delta^{(1)} S^{(3)}+\delta^{(2)} S^{(2)} =0,
\end{equation}
which  involves quartic vertices $S^{(4)}$. 

A conclusive analysis of higher-spin interactions in the covariant form at this  quartic order was 
not performed so far, 
but to fix $g_{0s_2s_3}$  we may   use  the result of  Metsaev 
 \cite{Metsaev:1991mt} obtained  in  $d=4$ in the light-cone gauge 
approach.
Making  the most general  ansatz for the cubic interaction vertex  and  requiring the closure of the Poincare algebra 
 to the $g^2$ order it was   found   \cite{Metsaev:1991mt}  
   that all cubic couplings can be expressed in terms of
 a  single   parameter.
  In Appendix A we  will  establish a dictionary between the  light-cone  cubic vertices
 and the covariant ones  in the de Donder gauge and show that the result
  of \cite{Metsaev:1991mt,Metsaev:1991nb} when translated into the covariant
 language implies that 
\begin{equation}
\label{29}
g_{0s_2s_3}=g \frac{\ell^{s_2+s_3-1}}{(s_2+s_3-1)!}\ .
\end{equation}
Here  $g$  is  an  overall   dimensionless coupling counting the power of fields  and 
$\ell$ is a  unique dimensional  parameter (cf. \rf{1}). 
Here $g_{000}=0$, i.e. there is   no cubic scalar self-coupling. 

 Remarkably, the same expression for the cubic 
  couplings  \rf{29}  appears also in the action \rf{2}  for the  massless   higher spin fields  in AdS$_4$  which was 
  reconstructed from the condition of  consistency with the vectorial AdS/CFT  duality. 
Namely, in \cite{Skvortsov:2015pea} it was noted that $g_{00s}$ couplings   \rf{29}  agree with 
their   AdS counterparts reconstructed from the free  boundary  CFT \cite{Bekaert:2015tva}
and it was  further conjectured that the agreement should hold  also for general trilinear 
couplings  of higher-spin gauge fields. This was indeed confirmed recently in 
 \cite{Sleight:2016dba}.
  This gives  hope that 
  some relation between the AdS action  \rf{2}  and the  flat space  one \rf{1}  may hold 
  even beyond the cubic order (despite a  naive flat-space limit of \rf{2} being singular).\foot{Let us also mention that instead 
of solving a full-fledged version of (\ref{nquartic}) one  may  study
a simpler consequence of it:  one may  consider  deformations of gauge transformations 
induced by  a  cubic  vertex  and demand that they  satisfy a generalised version
 of the Jacobi identity \cite{Berends:1984rq,Barnich:1993vg}. 
 In   higher-spin theory in AdS  it was shown  \cite{Fradkin:1986ka,Boulanger:2013zza}
 that under some  natural assumptions  the general solution to the Jacobi identity  is 
 quite  constraining.
Then similarly to (\ref{29}) this   should allow one  to express  all non-Abelian cubic couplings  in terms of a single
 coupling constant $\ll$, as was shown in the light-cone gauge in  \ci{Metsaev:1991mt,Metsaev:1991nb}.
 }

\iffa 
 It is said that there is one-parameter family of HS algebras. It is confusing. 
 There is a one-parameter family relevant for AdS_5, but this is a very special case.
  In Konstein-Vasiliev they studied the most general case of SUSY with
   YM groups, so there are many parameters.
    Perhaps it was meant that there is a single coupling constant in the action. 
\fi


\subsection{Deformation of gauge transformations}

In addition to cubic interactions of the physical fields we need also to find  similar vertices involving the ghost fields
as these are required to compute the one-loop self-energy graphs. 
To find the ghost action  corresponding to 
 the de Donder gauge  we need  to know  %
the  deformation of the free gauge transformations \rf{3}
 induced by  the  presence of the cubic  vertex (\ref{8}).\foot{Deformations of gauge transformations induced by cubic couplings of higher spin fields and the 
associated gauge algebra deformations have been studied, in particular, in
\cite{Berends:1984rq,Bekaert:2005jf,Boulanger:2006gr,Bekaert:2010hp,Joung:2013nma}.
However, due to certain issues (such as possible non-trivial contributions  from 
non traceless-transverse terms, ambiguity in field redefinitions, etc.) these results cannot be    directly
used   here. 
}

This   deformation may be found  using  (\ref{11}). 
In order to compensate for the term containing $\partial_{x_3}^2$ in (\ref{9}) one has to deform
 the  spin $s_3$  field gauge transformation.
In general,  one has
\begin{equation}
\label{12}
\delta^{(1)}_{{s_2}}\phi_{s_3}\,  \frac{\delta S^{(2)}[\phi_{s_3}]}{\delta \phi_{s_3}} = s_3! \int d^dx \left[ \delta^{(1)}_{{s_2}} \phi_{s_3}(x_3,\partial_{u_3})\,\hat T_3\,
\partial^2_{x_3} \phi_{s_3}(x_3,u_3)\right]_{u_3=0}\ .
\end{equation}
Substituting (\ref{9}) and (\ref{12}) into (\ref{11}) we find
\begin{align}
\notag
\int d^dx \Big[-g_{0s_2s_3}  s_2&(\partial_{u_2}  \cdot \partial_{x_{31}})^{s_2-1}
(\partial_{u_3} \cdot \partial_{x_{12}})^{s_3}
  \phi_0(x_1)\varepsilon_{s_2-1}(x_2,u_2)\partial^2_{x_3}\phi_{s_3}(x_3,u_3)\\
  \label{12proc1}
  & \qquad\qquad\qquad\quad+
  s_3!\;\delta^{(1)}_{s_2} \phi_{s_3}(x_3,\partial_{u_3})\,\hat T_3\,
\partial^2_{x_3} \phi_{s_3}(x_3,u_3)
  \Big]_{\substack{{u_i=0}\\{x_i=x}}}=0.
\end{align}
Employing (\ref{12proc2})  we get
\begin{align}
\notag
\int d^dx \Big[ \Big( s_3! \;\delta^{(1)}_{s_2}& \phi_{s_3}(x_3,\partial_{u_3})-g_{0s_2s_3}  s_2(\partial_{u_2}  \cdot \partial_{x_{31}})^{s_2-1}
(\partial_{u_3} \cdot \partial_{x_{12}})^{s_3} \, \hat T^{-1}_3\,   \phi_0(x_1)\,\varepsilon_{s_2-1}(x_2,u_2) 
\Big)
\\
  \label{12proc3}
& \qquad \quad\times  
 \,\hat T_3\,
\partial^2_{x_3} \phi_{s_3}(x_3,u_3)
  \Big]_{\substack{{u_i=0}\\{x_i=x}}}=0\ .
\end{align}
After some simple manipulations this yields
\begin{align}
\notag
\delta^{(1)}_{s_2} &\phi_{s_3}(x_3,u_3) =  g_{0s_2s_3}\frac{s_2}{s_3!}\Big[ \Big\{\Big\{ \,\hat T^{-1}_3\, ({u_3}\cdot \partial_{x_{12}})^{s_3}
 \big(\partial_{u_2}\cdot (-2\partial_{x_1}-\partial_{x_2})\big)^{s_2-1}\\
\label{14}
& \qquad\qquad\qquad\qquad\qquad \qquad\times\phi_0(x_1)\,\varepsilon_{s_2-1}(x_2,u_2)\Big\}\Big\}
\Big]_{x_1=x_2=x_3},
\end{align}
where $\{\{... 
\}\}$ denotes the double-traceless projection in $u_3$, which in (\ref{12proc3}) 
was enforced  implicitly by contraction
with double-traceless $\phi_{s_3}$.

The presence of $\partial_{x_1}^2$ in (\ref{9}) indicates
 that the cubic vertex  (\ref{8}) 
also induces  a non-trivial  transformation of the spin 0  field (which was not transforming at the leading order
in \rf{3}) 
with respect to  the gauge  symmetries of the  
 higher spin fields. 
 The explicit  knowledge of this  transformation will not, however,  be required 
 for  the  computation of the one-loop scalar self-energy below. 
 
  Let us  remark on an  ambiguity in the  cubic vertex related to field redefinition freedom. 
 The perturbative deformation procedure of the free gauge theory always has an  ambiguity of    additional 
 local field redefinitions. 
 Some preferred choice of   basic fields    may be selected  by  the requirement of the 
 most simple  form of the  full non-linear theory consistent with  manifest symmetries  (cf. Einstein theory). 
  Here we   shall choose the  ``minimal"  form (\ref{8}) of the   cubic   vertex which is universal for all spins.
 When solving the gauge invariance condition (\ref{11}) one usually 
ignores  cubic vertices that vanish on the free equations of motion. 
Such vertices are ``fake"  in the sense
that they can be generated from the free theory Lagrangian by field redefinitions. 
For example, in addition to the non-trivial vertex (\ref{8})  one may  consider, e.g., for  $s_2=s_3=s$
\begin{equation}
\label{df}
S^{(3)}[\phi_0,\phi_{s},\phi_{s}] \sim \int d^dx \,\phi_0(x)\; {\cal R}_s (x)\cdot {\cal R}_s(x)\ ,
\end{equation}
 where ${\cal R}_s$ is the de Wit - Freedman curvature \cite{deWit:1979sib}  which is manifestly invariant under the linearized gauge transformations \rf{3} (and thus 
does not induce any gauge symmetry deformation,  i.e. $\delta^{(1)}\phi_s =0$).
This does not contradict  (\ref{14}) implying
  that the deformation of the gauge symmetry
 is non-zero for $s_2=s_3$.
 Indeed, one can show that  \rf{14} for $s_2=s_3$  can be removed by   a field redefinition.\foot{For different spins $s_2 < s_3$ the deformation of the gauge symmetry of the spin $s_3$ field is always non-trivial
 \cite{Bekaert:2010hp,Joung:2013nma} and  
 the vertex cannot be put into the form
 (\ref{df}).} 
 
 The local  field  redefinitions that preserve the asymptotic states 
  should not,  of course,  change the on-shell amplitudes. 
  That means that the  ``contact"   contribution of the ``one-shell trivial" 
  3-point vertex containing the free kinetic operator acting on one of the legs 
  (i.e. the  one that  can be removed by a field redefinition) to, e.g., the 4-point scattering amplitude  will  be cancelled 
  by the  contribution of the 4-vertex   produced by the field redefinition.

 \iffa
 Let us also comment on the effect of field redefinitions on the on-shell amplitudes.
 For definiteness, we consider an on-free-shell trivial cubic interaction. As argued above,
  this can be removed by a field
 redefinition. This redefinition, however, generates a quartic vertex. For the
 tree-level  four-point amplitude the effect of such a field redefinition amounts to the replacement of
 the exchange diagram by the contact one. One can show that the amplitude, nevertheless, remains the same.
 More generally, one can show that field redefinitions do not change total on-shell amplitudes.
 They may change, however, contributions of individual Feynman diagrams.
 \fi


\subsection{Ghost action}

According to the standard Faddeev-Popov procedure, 
the  ghost  action  for the free   theory  \rf{1},\rf{5} invariant under \rf{3}  in the de Donder gauge \rf{4}  is 
\begin{align}
\notag
&S^{(2)}[\bar c_{s-1},c_{s-1}]= (s-1)! \int d^dx\Big[ \bar c_{s-1}(x,\partial_u) 
\,\hat D\,
\frac{\delta^{(0)}_{s} \phi_s}{\delta \varepsilon_{s-1}}
 c_{s-1} (x,u)\Big]_{u=0}\\
 \label{15}
 &\qquad\qquad\qquad\qquad\qquad\qquad\qquad=(s-1)! \int d^dx \Big[\bar c_{s-1}(x,\partial_u) \partial_x^2 c_{s-1} (x,u)\Big]_{u=0}\ .
\end{align}
The cubic vertex involving the ghost fields is  found analogously by 
using  the  deformation of the gauge transformation (\ref{14})
instead of the leading-order transformation $\delta^{(0)}_s \phi_s$ in \rf{15} 
\begin{align}
\notag
&S^{(3)}[\bar c_{s_3-1},c_{s_2-1},\phi_0]
= (s_3-1)! \int d^dx \Big[\bar c_{s_3-1}(x_3,\partial_{u_3}) 
 \,\hat D_3\,
\frac{\delta^{(1)}_{s_2} \phi_{s_3}}{\delta \varepsilon_{s_2-1}}
 c_{s_2-1} (x_3,u_3)\Big]_{u_3=0}\\
\notag
 &=-\;g_{0s_2s_3}\, \frac{s_2}{s_3} \int d^dx \Big[\bar c_{s_3-1}(x_3,\partial_{u_3})
\Big((\partial_{x_1}+\partial_{x_2}) \cdot \partial_{u_3}-\ha
u_3\cdot (\partial_{x_1}+\partial_{x_2})\; \partial^2_{u_3}  \Big)\\
&  \qquad  \times   \Big\{\Big\{ \,\hat T^{-1}_3\, 
 ({u_3}\cdot \partial_{x_{12}})^{s_3}
 \Big(\partial_{u_2}\cdot (-2\partial_{x_1}-\partial_{x_2})\Big)^{s_2-1}
    \phi_0(x_1)
 c_{s_2-1} (x_2,u_2)\Big\}\Big\}\Big]_{\substack{{u_i=0}\\{x_i=x}}}\ .
  \label{15new}
\end{align}
The double-traceless projector in $u_3$ can be dropped as 
it is already  imposed by the  contraction with the remaining 
part of the integrand.
After some straightforward algebra  eq.(\ref{15new}) acquires a remarkably simple form
\begin{align}
\notag
S^{(3)}[\phi_0,\bar c_{s_3-1}, c_{s_2-1}] &=-\; g_{0s_2s_3}\, s_2 \int d^dx \Big[ (\partial_{u_3}\cdot \partial_{x_{12}})^{s_3-1} 
(\partial_{u_2} \cdot \partial_{x_{31}})^{s_2-1} (\partial_{x_{12}}\cdot \partial_{x_3})\\
\label{18}
& \qquad\qquad \qquad  \qquad \times 
\bar c_{s_3-1}(x_3,u_3) \phi_0(x_1) c_{s_2-1}(x_2,u_2) \Big]_{\substack{{u_i=0}\\{x_i=x}}}\ . 
\end{align}

\section{Feynman rules}

The  propagator for the free  Fronsdal   field   in the de Donder gauge 
was originally found in $d=4$ in \cite{Fronsdal:1978rb} and was later extended to  any  dimension $d$ 
in \cite{Francia:2007qt}. For completeness,  we  shall review its  derivation  below. 
The expression for  the propagator in terms of the Gegenbauer (or Chebyshev in $d=4$) polynomials   was   given in \ci{Bekaert:2009ud}  that we follow here.

It is convenient to split the double-traceless  field $\p_s$
into two traceless fields $\nu_s$  and $\mu_{s-2}$   as
\begin{equation}
\label{19}
\phi_s(x,u) = \nu_s (x,u)+ \frac{1}{s(s-1)} u^2 \mu_{s-2} (x,u)\ .
\end{equation}
Then  the gauge fixed action (\ref{5})  becomes 
\begin{align}
\notag
S^{(2)}[\phi_s]&=\frac{s!}{2}\int  d^d x \, \left[\; \nu_s(x,\partial_u) \partial_x^2\;
  \nu_s(x,u)\right]_{u=0}\\
  \label{19action}
   &-\frac{(d+2s-4)(d+2s-6)}{s(s-1)}\, \frac{(s-2)!}{2}
   \int  d^d x \, \left[ \mu_{s-2}(x,\partial_u) \partial_x^2\;
  \mu_{s-2}(x,u)\right]_{u=0}\ .
\end{align}
The Fronsdal  field propagator is found to be 
\begin{equation}
\label{Frprop}
{\cal D}_s^d(u,u';p) = -\frac{i}{p^2}\, \Big[ {\cal P}_s^d (u,u')+\frac{s(s-1)}{(d+2s-4)(d+2s-6)}\, \frac{u^2 u'^2}{s^2(s-1)^2}\,  {\cal P}_{s-2}^{d} (u,u')\Big]\ ,
\end{equation}
where ${\cal P}_s^d$ is a generating function for the standard rank-$s$ traceless projector bi-tensor
\begin{equation}\la{08}
{\cal P}_s^d(u,u') = \frac{1}{(s!)^2}(u\cdot u')^s +\dots, \qquad \qquad \partial_u^2 \,{\cal P}_s^d(u,u') =\partial_{u'}^2{\cal P}_s^d(u,u') =0\ .
\end{equation}
It can be defined as a series
\begin{equation}
\label{20}
{\cal P}^{d}_s(u,u') =\frac{1}{(s!)^2}\sum_{k=0}^{[s/2]}
t^d_{s,k}   (u^2)^k (u'^2)^k (u\cdot u')^{s-2k},
\end{equation}
where 
\begin{equation}
\label{21}
t^d_{s,k}= \frac{(-1)^k s!}{4^k k! (s-2k)! ({d\ov 2}-1+s-k )_k}
\end{equation}
and $(a)_k = \Gamma(a+k)/\Gamma(a)$   is the Pochhammer symbol.
It is  convenient to rewrite  (\ref{20})  as
\begin{equation}
\label{20G}
{\cal P}^d_{s}(u,u')=\frac{1}{(s!)^2} \frac{s!}{ ({d\ov 2}-1)_s} \left(\ha {\sqrt{u^2 u'^2}}\right)^s C_s^{\frac{d}{2}-1} 
\Big(\frac{u\cdot u'}{\sqrt{u^2 u'^2}}\Big)\ ,
\end{equation}
where
\begin{equation}\la{010}
C_s^\alpha (z) \equiv \sum_{k=0}^{[s/2]} \frac{(-1)^k (\alpha)_{s-k}}{k! (s-2k)!} (2z)^{s-2k}
\end{equation}
is the Gegenbauer 
polynomial.
Observing   that the coefficients in \rf{21} satisfy 
\begin{equation}
t^d_{s,k} - \frac{s(s-1)}{(d+2s-4)(d+2s-6)}t^{d}_{s-2,k-1}= t^{d-2}_{s,k}\ , 
\end{equation}
or using the identity 
\begin{equation}
C^{\alpha}_s(z) - C_{s-2}^{\alpha}(z) = \frac{\alpha+s-1}{\alpha-1}C_s^{\alpha-1}(z)\ , 
\end{equation}
the propagator (\ref{Frprop}) can be  put into  the following simple form 
\begin{equation}
\label{Frprop1}
{\cal D}_s^d(u,u';p) = -\frac{i}{p^2}\, {\cal P}_s^{d-2} (u,u')\ .
\end{equation}
Thus, in  agreement with the result of  \cite{Francia:2007qt},  the tensorial part of
the Fronsdal propagator is just the traceless projector in $d-2$ dimensions.

Let us  note that for $d=4$ the higher spin propagator  contains  the  two-dimensional
projector ${\cal P}^2_s$, for which the representation in terms of the Gegenbauer polynomial
(\ref{20G})  is singular  due to vanishing of the denominator in the prefactor. 
Instead, one can use the expression 
\begin{equation}
\label{20Gd4}
{\cal P}^2_{s}(u,u')=\frac{1}{(s!)^2} 2 \left(\ha {\sqrt{u^2 u'^2}}\right)^s T_s 
\Big(\frac{u\cdot u'}{\sqrt{u^2 u'^2}}\Big)\ ,
\end{equation}
where 
\begin{equation}
\label{cheb}
T_s (z) \equiv \frac{s}{2}\sum_{k=0}^{[s/2]} \frac{(-1)^k (s-k-1)!}{k! (s-2k)!} (2z)^{s-2k}
\end{equation}
is the Chebyshev polynomial of the first kind.
 It can also be defined by
  \begin{equation}
  \label{chebid}
 T_s(z)=\ha\Big[\big(z+\sqrt{z^2-1}\big)^s+\big(z-\sqrt{z^2-1}\big)^s\Big]\ .
 \end{equation}
The free ghost field in (\ref{15})    is canonically normalised and traceless, i.e.  has   propagator 
\begin{equation}
\label{24?}
{\cal G}^{d}_{s-1}(u,u';p) = -\frac{i}{p^2}\, {\cal P}^{d}_{s-1}(u,u')\ .
\end{equation}

The cubic vertex for physical fields can be easily obtained from  (\ref{8})
\begin{equation}
\label{416}
{\cal V}(\partial_{u_2},\partial_{u_3};p_1,p_2,p_3) = 2 ig_{0s_2s_3} (-ip_{31} \cdot \partial_{u_2})^{s_2} (-ip_{12}\cdot \partial_{u_3})^{s_3},
\end{equation}
where 
$p_{ij}\equiv p_i - p_j$
and $2$ is a symmetry factor.\footnote{Such symmetry factor appears when the vertex contains 
two identical fields. In the considered case  spins $s_2$ and $s_3$ may be different. However,
eventually, we are going to sum over spins and both $\phi_{s_2}$ and $\phi_{s_3}$ appear
as particular members of two identical higher spin multiplets. Alternatively, this factor
can be understood by arguing that during summation over spins each unordered pair $(s_2,s_3)$ is counted 
twice.}
 This vertex is non-trivial only when $s_2+s_3$ is even (\ref{totalodd}).
Similarly,  for the ghost vertex we find from (\ref{18})
\begin{equation}
\label{417}
{\cal W}(\partial_{u_2},\partial_{u_3};p_1,p_2,p_3) = ig_{0s_2s_3} s_2 (-ip_{31}\cdot \partial_{u_2})^{s_2-1} (-ip_{12}\cdot\partial_{u_3})^{s_3-1} (-p_{12}\cdot p_3)\ .
\end{equation}

The  resulting  set of Feynman rules for the physical  fields (wavy line or solid line for $s=0$)   and the ghosts  (dotted  line) 
 can be summarised as follows
\begin{eqnarray}
\begin{aligned}
\begin{fmffile}{fpropagator}
\begin{fmfgraph}(20,5)
\fmfleft{i1}
\fmfright{o1}
\fmf{photon}{i1,o1}
\end{fmfgraph}
\end{fmffile}\no 
\end{aligned}
&\qquad = \qquad &
{\cal D}_s^d(u,u';p)
\\
\begin{aligned}
\begin{fmffile}{pvertex}
\begin{fmfgraph}(20,20)
\fmfleft{v1,v4}
\fmfright{v2}
\fmf{photon}{v1,v3}
\fmf{photon}{v4,v3}
\fmf{plain}{v3,v2}
\fmfdot{v3}
\end{fmfgraph}
\end{fmffile}
\end{aligned}
&\qquad = \qquad&
{\cal V}(\partial_{u_2},\partial_{u_3};p_1,p_2,p_3)
\label{24}
\\
\begin{aligned}
\begin{fmffile}{gpropagator}
\begin{fmfgraph}(20,5)
\fmfleft{i1}
\fmfright{o1}
\fmf{ghost}{i1,o1}
\end{fmfgraph}
\end{fmffile}\no 
\end{aligned}
&\qquad = \qquad&
{\cal G}^{d}_{s-1}(u,u';p)
\\
\begin{aligned}
\begin{fmffile}{gvertex}
\begin{fmfgraph}(20,20)
\fmfleft{i1,i2}
\fmfright{v2}
\fmf{plain}{v1,v2}
\fmf{ghost}{i1,v1,i2}
\fmfdot{v1}
\end{fmfgraph}
\end{fmffile}
\end{aligned}& \qquad = \qquad &
{\cal W}(\partial_{u_2},\partial_{u_3};p_1,p_2,p_3) 
\no 
\end{eqnarray}
where explicit expressions are given in (\ref{20G}),\rf{20Gd4},(\ref{Frprop1}), (\ref{24?})-(\ref{417}) 
and the momenta are assumed to be ingoing.

\def \s   {{\rm s}} \def \t {{\rm t}}   \def \u {{\rm u}}

 \section{Tree-level 4-scalar scattering  amplitude} 
 
 In this section we shall compute the tree-level  4-scalar scattering    amplitude 
  due to exchange of  the  tower of higher spin fields. 
 For charged     external scalar particles and arbitrary coupling constants in the corresponding cubic vertex 
 this  was earlier  discussed  in \cite{Bekaert:2009ud}. 
 Here we shall  repeat the same computation in  the case
 of a  real scalar which is the $s=0$  member  of the tower of higher spin fields 
  using the  specific values of the  cubic couplings  given by  (\ref{29}).\foot{
  While   our final expression  may be viewed  as a special  case of the  general one   in 
  \cite{Bekaert:2009ud}
  the use of the particular  values of the couplings in \rf{29} is important as it  leads  to a convergent 
  sum over all  higher spin contributions   and thus  to  a completely  explicit 
  expression for the exchange amplitude.
  }
 
 The important difference between the present case of a real scalar  scattering and a complex scalar one in 
 \cite{Bekaert:2009ud} is that  in the real scalar case   all   interactions with odd spins vanish  (cf. \rf{8}).
 In the   complex scalar  case   
the   odd spins contribute to the
exchange amplitude  with the opposite sign compared to the even spins. The  resulting sum over spins 
contains terms with alternating signs 
 which leads to an improved high energy behaviour \cite{Bekaert:2009ud}
 compared  to the real scalar scattering  case, where all exchange  contributions
add up with the same sign.\foot{Similar 
softening mechanism was discussed, e.g.,  in \cite{GSW}.} 
Note that this  remark  applies  to the full amplitude  assuming the ``contact" 4-point  vertex  contribution does not alter 
the UV asymptotics of the amplitude which need not be the case  here (see   discussion below). 
 
  \subsection{Exchange contribution}
  
 The  $\s$-channel exchange of spin $j$  field  gives  (see \rf{24})
 \begin{equation}\la{51}
 \begin{aligned}
\begin{fmffile}{exchangesmall}
\begin{fmfgraph}(20,15)
\fmfleft{i1,i2}
\fmfright{o1,o2}
\fmf{plain}{i1,v1,i2}
\fmf{plain}{o1,v2,o2}
\fmf{photon}{v1,v2}
\fmfdot{v1}
\fmfdot{v2}
\end{fmfgraph}
\end{fmffile}
\end{aligned} \equiv \quad{\cal A}^j_{exch}(\s,\t,\u) \quad =\quad {\cal V}(\partial_u;p_1,p_2,p){\cal V}(\partial_{u'}; p'_1,p'_2,p){\cal D}_j^{d}(u,u';p)\ ,
 \end{equation}
 where all the external momenta $p_1$, $p_2$, $p'_1$ and $p'_2$ are assumed to be  ingoing: $p_1$ and $p_2$
 into the left vertex and $p_1'$ and $p_2'$ into the right one  ($p=p_1+p_2=-p_1'-p_2'$). 
  Let us  introduce the Mandelstam
 variables 
 \begin{equation}\la{511}
 \s\equiv -(p_1+p_2)^2, \quad \t\equiv-(p_1+p_1')^2, \quad \u\equiv -(p_1+p'_2)^2\ , \quad 
 \s+\t + \u=0 \ , \ \ \ p_i^2=p_i'^2=0 \ . 
 \end{equation}
 Using (\ref{Frprop1}), (\ref{416}) and that 
 $p^2_{12} = p'^2_{12}=-\t-\u, \quad p_{12}\cdot p'_{12} = -\t+\u$
  we find\foot{Here  the symmetry factor  2 in the  cubic  vertices  \rf{416}  is  not needed.} 
 \begin{equation}
{\cal A}_{exch}^j(\s,\t,\u)= -\frac{i g^2_{00j}}{\s}\frac{ 2^{-j} j! }{({d\ov 2}-2)_j} \big({t+u}\big)^j\, C_j^{\frac{d}{2}-2}
\te \big(\frac{\t-\u}{\t+\u}\big)\ .   \la{52}
 \end{equation}
 In $d=4$ for an individual spin $j$ contribution we  obtain
   \begin{equation}\la{53}
 {\cal A}_{exch}^j(\s,\t,\u) = -\frac{i g^2_{00j}}{\s}  2^{-j+1}  \left({\t+\u}\right)^{j} \;T_{j}
\te \big(\frac{\t-\u}{\t+\u}\big)\ , 
 \end{equation}
 where $T_j(z)$ is  given in (\ref{cheb}),\rf{chebid}.
  
 Summing over all  even spins $j=2k$ we obtain in $d=4$
    \begin{equation}\la{54} 
 {\cal A}_{exch}(\s,\t,\u) = \sum_{j=0,2,4,...}^{\infty} {\cal A}_{exch}^{j}(\s,\t) =   -\frac i \s\sum_{k=0}^{\infty}{ g^2_{00\,2k}}\,  2^{-2k+1}  \left({\t+\u}\right)^{2k} \;T_{2k}
\te \big(\frac{\t-\u}{\t+\u}\big)\ . 
 \end{equation}
 Let us introduce the function 
 \begin{equation}\la{55}
 F(z)\equiv \sum_{k=0}^{\infty} g^2_{00\,2k}\, z^{2k} =g^2 \sum_{k=0}^{\infty} \frac{1}{\big[(2k-1)!\big]^2}\, ( \ll^2 z)^{2k}
 \ , 
 \end{equation}
 where   we used the values of the couplings  in \rf{29} (note that here the $k=0$ term vanishes). 
 It can be expressed in terms of the Bessel  and the modified Bessel functions
\begin{align}\la{5.7}
J_\alpha(z) &= {\te  \left(\frac{z}{2}\right)^\alpha} \sum_{k=0}^\infty \te 
\frac{(-1)^k}{\Gamma(k+1)\Gamma(\alpha+1+k)}\left(\frac{z}{2}\right)^{2k}\ , \quad
I_\alpha(z) &={ \te  \left(\frac{z}{2}\right)^\alpha} \sum_{k=0}^\infty\te 
\frac{1}{\Gamma(k+1)\Gamma(\alpha+1+k)}\left(\frac{z}{2}\right)^{2k}
\end{align}
as  follows 
\begin{equation}\la{56}
 F(z)
= \ha g^2 \, {\ell^2z} \Big(I_0(2\ell\sqrt{z})-J_0(2\ell\sqrt{z})\Big)\ .
\end{equation}
Using (\ref{chebid})
 the  $\s$-channel amplitude \rf{54}   in $d=4$   may  then be written as\foot{In contrast  to a similar amplitude in conformal higher-spin theory 
 \ci{Joung:2015eny}   here the sum  over spins   is convergent and the amplitude has 
 non-trivial (non-scale-invariant)  dependence on the  Mandelstam variables   due to the presence of the 
 dimensionful parameter  $\ll$.}
 \begin{equation}\la{57}
 {\cal A}_{exch}(\s,\t,\u)= -\frac{i}{\s}\Big[ F\Big(\ha{(\sqrt{\s+\t}+\sqrt{ \t}\, )^2}\Big) + F\Big(\ha{(\sqrt{\s+\t}-\sqrt{\t}\, )^2}\Big) 
  \Big] \ .
 \end{equation}
 Note that  for $\s\to 0$  the $1\ov \s$   pole  is not cancelled 
 \be
 \la{777}
\s\to 0 :  \qquad  {\cal A}_{exch}(\s,\t,\u)= -\frac{i}{\s}\Big[ g^2 \ell^2  \t   \Big(I_0(2\ell\sqrt{2 \t})-J_0(2\ell\sqrt{2\t})\Big) + {\cal O} (\s) 
  \Big] \ .
  \ee
 The full exchange  amplitude is found by adding  also  the   contributions of the $\t$ and $\u$ channels, 
 \be \la{577}
  \hat {\cal A}_{exch}(\s,\t,\u)=  {\cal A}_{exch}(\s,\t,\u) + {\cal A}_{exch}(\t,\s,\u) + {\cal A}_{exch}(\u,\t,\s) \ . 
  \ee
In the Regge limit ($\t\to\infty$, \ $\s$=fixed)  
the  $\s$-channel   amplitude \rf{57}  gives the   dominant   contribution 
and using 
the standard  asymptotics 
\be \la{588}  z\to \infty: \qquad \te 
I_\alpha(z) = {\frac{1}{\sqrt{2\pi z}}}\, e^z + ...\ , \ \  \qquad 
 \  J_\alpha(z) = \sqrt{\frac{2}{\pi z}} \cos{\left(z-\frac{\alpha\pi}{2}-\frac{\pi}{4}\right)} + ... \ , 
 \ee
 we find 
\begin{equation}\la{58} \t\to\infty, \ \s={\rm fixed} : \quad  
\hat {\cal A}_{exch}(\s,\t,\u)\sim -\frac{i g^2}{\s} \ell^2 \t\, I_0(2\ell\sqrt{2\t})\sim  -\frac{i g^2}{\s}  \frac{ (\ll^2\t)^{3/4}}{2^{5/4}\pi^{1/2}} 
\, e^{2\ll\sqrt{2\t}}
\end{equation}
Similar exponentially growing expression  is found also in the fixed-angle limit (cf.  \cite{Bekaert:2009ud}). This  contrasts   with soft UV behaviour   of string scattering amplitudes.
Analogous results  are expected also for other spin $s$  scattering amplitudes. 
Such   exponential growth of the tree-level scattering amplitude in the high energy regime is   an indication of the presence of 
similar ultraviolet divergences in loop diagrams.  Indeed, we shall find similar  UV divergences in the  one-loop 
self-energy  bubble  diagram contribution  discussed in the next   section. 

Let us note that despite $\sqrt \s$ and $\sqrt \t$   arguments  appearing  in  
 \rf{57} the exchange amplitude  is actually analytic in $\s,\t$. 
Indeed, the Bessel functions in \rf{56} have expansion in even powers of their  arguments  and,  as a consequence, 
the sum of the two $F$-functions  in \rf{57} has a convergent  expansion in integer powers of $\s$ and $\t$ only. 
The  appearance of a non-analytic  function $e^{a\sqrt{\t}}$ in \rf{58}  is  just   an artifact of the large $\t$ limit.\foot{To give a simple example, $\cosh \sqrt z$ is an analytic function on the $z$-plane (given by  a series  of  integer  powers of $z$) 
but its large $z$  asymptotics  is the  same as   that of  a non-analytic  function $\ha e^{\sqrt z}$.}  

\subsection{Comments on 4-vertex  contribution}

 The  full tree-level 4-point spin 0  scattering amplitude  should  also include   the 
contribution of the 4-point   0-0-0-0  vertex: 
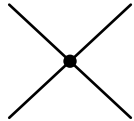
\be\no
\begin{aligned}
\begin{fmffile}{contactsmall}
\begin{fmfgraph}(20,15)
\fmfleft{i1,i2}
\fmfright{o1,o2}
\fmf{plain}{i1,v1}
\fmf{plain}{i2,v1}
\fmf{plain}{o1,v1}
\fmf{plain}{o2,v1}
\fmfdot{v1}
\end{fmfgraph}
\end{fmffile}
\end{aligned}
\ee
This vertex  is expected to  be effectively non-local (i.e.  may contain  an infinite series of  powers of derivatives) 
and may thus  ``soften"   the  large momentum  behaviour of the exchange  diagram contribution. 

While the precise form of the  4-scalar vertex in the flat-space   action \rf{1}  is  presently  unknown, we may 
  try to  get some idea about its structure 
   using an     analogy with  a similar term  in the action \rf{2}   of 
  the massless higher-spin theory in AdS. The 0-0-0-0  term in the AdS  action  was recently reconstructed 
  \cite{Bekaert:2015tva} from the  free boundary CFT data   by assuming the    AdS/CFT  correspondence. 
We shall use a heuristic trick of 
replacing the AdS covariant derivatives by  the flat space   ones  in the expression in    \cite{Bekaert:2015tva}. This then   gives
\be\la{59}
S^{(4)}[\phi_0] = g^2 \int d^4x \Big[
\sum_{j=0}^\infty f_{2j}\big(\Delta_{x_{34}}\big)\,  \big(\partial_{x_{12}}\cdot  \partial_{x_{34}}  \big)^{2j}
\,   \phi_0(x_1)\phi_0(x_2)\phi_{0}(x_3)\phi_0(x_4)\Big]_{x_i=x}\ ,
\ee
where $\Delta_{x_{34}} \equiv  (\partial_{x_{3}} + \partial_{x_{4}})^2, \ \ \partial_{x_{12}} \equiv  \partial_{x_{1}} - \partial_{x_{2}}$
and the function $f_{2j}(z)$   is given by some infinite power  series in $z$
and is regular at $z\to 0$. 
To estimate the  large momentum expansion  of the resulting amplitude we may choose  the large $z$  asymptotics  of 
 the function  $f_{2j}(z)$ to be the same as in 
 corresponding function in the   AdS   4-scalar vertex    in  \cite{Bekaert:2015tva}, i.e. 
 \begin{equation}\la{60}
 f_{2j} (z) =  c_{2j}\frac{ \ll^{4j-2}}{z}  \ , \ \ \ \ \ \  \qquad    z\to \infty \ . 
 \end{equation}
Here  $c_{2j}$ are numerical coefficients   and we introduced the parameter  $\ll$ to balance the dimensions.\foot{Note that 
the exact  expression for   $f_{2j} (z)$  in   \cite{Bekaert:2015tva}   does not have  poles (in particular, is regular  at $z\to 0$), so 
its flat-space counterpart  should also    not contain an essential non-locality  like $1\ov \del^2$.}
 Given   \rf{59},\rf{60}  the contribution of the 4-vertex
 to the 4-scalar amplitude  
 may, in principle,  cancel the exponential growth \rf{58}  of the exchange amplitude.  
 For example, choosing $c_{2j}$  in the form   similar to the one  they  have in  the AdS expression
   \cite{Bekaert:2015tva}
 \be \la{61} 
 c_{2j}=  \frac{1}{
  [(2j-1)!]^2}\ ,\ee
 we  find  that the asymptotic  contribution  of the vertex \rf{59} to the 4-point amplitude is  proportional to 
  \begin{align}\no 
&\sum_{j=0}^\infty f_{2j}\big(-(p_3+p_4)^2\big)\, \big[(p_1-p_2)\cdot(p_3-p_4)\big]^{2j}=\sum_{j=0}^\infty  f_{2j}(\s)\, (\t-\u)^{2j}\\
&\qquad\qquad \qquad  =
 \frac{2\t + \s}{2\s} \left[I_0\big(2\ell\sqrt{2\t + \s}\big)-J_0\big(2\ell \sqrt{2\t + \s}\big) \right]\ , \la{62}
\end{align}
where $\s$ is assumed to be large.
The resulting contribution   is thus  also expressed in terms of the 
 Bessel functions  as in \rf{56},\rf{57}, i.e.   has  a similar Regge  behaviour as  in  \rf{58}. 


 \iffa 
  The reason for 
Up to a numerical prefactor this gives  the  dominant behaviour  at $z\to \infty$ of $a_i(z)$ found
in  \cite{Bekaert:2015tva} and $\ell$ is introduced on dimensional grounds.
 Then the tadpole contribution is
\begin{equation*}
\Theta(k^2)\sim \int \frac{d^4 p }{p^2} \sum_{j=0} \left[a_{2j}(- (k-p)^2) \left((k+p)^2\right)^{2j} + 
a_{2j} (-(k+p)^2) \left((k-p)^{2}\right)^{2j}\right].
\end{equation*}
 To estimate the UV behaviour of the integrand we neglect $k$ compared to $p$, which yields
 \begin{equation*}
 \Theta (0)\sim \int \frac{d^4 p}{p^2} \sum_{j=0}a_{2j}(-p^2) \left(p^{2} \right)^{2j}
 =
 \int \frac{d^4 p}{p^2}\, \frac{\ell^2 }{2}\, \frac{I_0(\sqrt{2}\ell p)-J_0(\sqrt{2} \ell p)}{2}.
 \end{equation*}
 We find that the tadpole diagram develops the same exponential behaviour in the UV
 as the bubble diagrams.
  So, taking into account tadpole
 diagrams can in principle make the self-energy UV finite.
 \fi

In the case of  higher-spin theory in AdS    one may    interpret the result for the 
tree-level amplitude   (that  reproduces the free CFT  4-point  correlator)  
as indicating a soft behaviour  in the UV. 
It was previously  observed  that the expressions  for the 
 Witten diagrams in AdS  written in the Mellin representation look  similar to the scattering amplitudes 
for the same processes in flat space  \cite{Penedones:2010ue}. 
In particular, the contact interactions with $2n$ derivatives 
give rise to the Mellin amplitudes given by  polynomials of $n$-th degree in the Mellin variables (which  play the role of 
the AdS counterparts of the Mandelstam variables). 
The AdS exchanges  produce  the Mellin amplitudes with poles  associated to the  dimensions
of the exchanged operators and their descendants. 
The  total four point scattering amplitude should  have a similar structure to 
the Mellin  amplitude   for  the  4-point correlator of spin 0 primary operator  in the free $O(N)$  CFT. This amplitude  is a 
distribution \cite{Taronna:2016ats,Bekaert:2016ezc} in the sense that
it is a certain combination of delta functions and hence it is zero everywhere except certain specific values of the Mellin variables. 
That could be interpreted as  suggesting 
 that the total tree-level   4-scalar  amplitude in flat space  may  also 
 be a  distribution and thus  should have   trivial UV asymptotics.\foot{Similar behaviour  was found 
  for the external scalar scattering amplitude in the  conformal
higher-spin theory \cite{Joung:2015eny}  (there the amplitude  actually vanished for the  physical
values of the Mandelstam variables).
}
It is   not clear, however, how such simple   amplitude could come out of   an  explicit flat-space  scattering computation   (for example,  adding the 4-vertex   contribution is  unlikely  to cancel the massless pole appearing in the exchange contribution).

Finally, let us  note  that the cubic vertices  (\ref{29})  appear to be  inconsistent with the BCFW constructibility
 condition. This was previously discussed in a similar context in \ci{Fotopoulos:2008ka,Dempster:2012vw}.
This condition  requires that the scattering  amplitudes vanish under infinite complex 
 shifts of momenta  \cite{Benincasa:2007xk}.
 Together with the assumption of analyticity\foot{If an analytic function vanishes at infinity it can be reconstructed  from poles
 and their residues.}    this leads to  certain 
recurrence relations  which  allow one  to express any tree-level amplitude in terms of the 
on-shell  three-point ones.
 Applied to the four-scalar   amplitude that we have studied   above 
 this would allow one  to  determine   the quartic scalar self-coupling in terms of the 
cubic vertices \rf{8},(\ref{29}). However, the BCFW  construction 
can be applied only if the cubic vertices satisfy a certain non-trivial consistency condition, the so 
called the four-particle test   \cite{Benincasa:2007xk}.
As we will show  in  Appendix B, the cubic vertices  \rf{8},(\ref{29}) fail to satisfy this test. 

It is not  clear   a  priori   if  
the condition of BCFW constructibility 
  should, in fact,  apply to an 
effectively non-local  theory like \rf{1}  containing  infinite number of  higher-spin fields  with higher derivatives  of any order in vertices. 
For example,  the BCFW  approach  relies on the assumption of the  analyticity  of the amplitudes. 
While perturbative amplitudes reconstructed from  higher spin  vertices will  be given by sums of integer powers of momenta 
(i.e.  are  analytic   before summation over spins)  the sums over spins could 
 not converge fast   enough.
This did not happen for  the exchange  amplitude \rf{57}   discussed above   but 
 was the case  for the 4-scalar scattering  in the conformal higher-spin theory  \ci{Joung:2015eny} 
 where the amplitude was not analytic -- was  given by a distribution.


\section{One-loop scalar self-energy  correction}

Let us now   consider the   spin 0  one-loop  self-energy correction.
 It  is given by  the sum of the  two  parts   -- the contribution of  bubble diagrams and 
 that  of tadpole diagrams. Each of these 
  contains contributions from the  physical  higher spin    loops  and from  the   ghost loops:
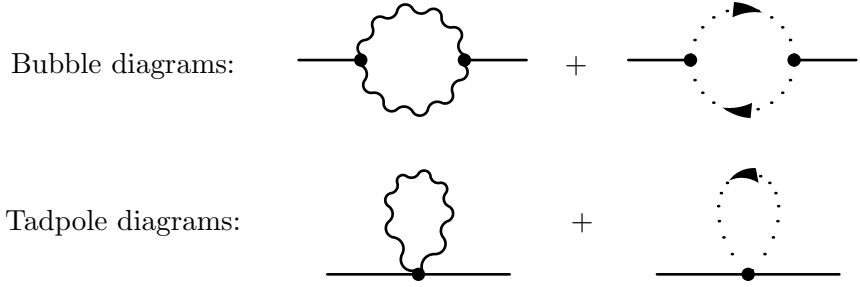
\begin{equation*}
\text{Bubble diagrams:} \qquad
\begin{aligned}
\begin{fmffile}{fselfenergy}
\begin{fmfgraph}(30,15)
\fmfleft{i}
\fmfright{o}
\fmf{plain}{i,v1}
\fmf{photon,left,tension=0.3}{v1,v2,v1}
\fmf{plain}{v2,o}
\fmfdot{v1}
\fmfdot{v2}
\end{fmfgraph}
\end{fmffile}
\end{aligned}\quad+ \quad
\begin{aligned}
\begin{fmffile}{gselfenergy}
\begin{fmfgraph}(30,15)
\fmfleft{i}
\fmfright{o}
\fmf{plain}{i,v1}
\fmf{ghost,left,tension=0.3}{v1,v2,v1}
\fmf{plain}{v2,o}
\fmfdot{v1}
\fmfdot{v2}
\end{fmfgraph}
\end{fmffile}
\end{aligned}
\end{equation*}
\begin{equation*}
\text{Tadpole diagrams:} \qquad
\begin{aligned}
\begin{fmffile}{ftadpole}
\begin{fmfgraph}(30,15)
\fmfleft{i1,i2}
\fmfright{o1,o2}
\fmf{plain}{i1,v1,o1}
\fmf{phantom}{i2,o2}
\fmf{photon,right,tension=0.6}{v1,v1}
\fmfdot{v1}
\end{fmfgraph}
\end{fmffile}
\end{aligned}\quad+ \quad
\begin{aligned}
\begin{fmffile}{gtadpole}
\begin{fmfgraph}(30,15)
\fmfleft{i1,i2}
\fmfright{o1,o2}
\fmf{plain}{i1,v1,o1}
\fmf{phantom}{i2,o2}
\fmf{ghost,right,tension=0.6}{v1,v1}
\fmfdot{v1}
\end{fmfgraph}
\end{fmffile}
\end{aligned}
\end{equation*}
The bubble diagrams are  straightforward  to evaluate    using the   Feynman   rules   described in section 4.
The tadpole diagrams   contain the  physical  4-point  0-0-$s$-$s$ vertices $S^{(4)}$ and  their ghost  counterparts 
(determined   by  subleading  gauge symmetry deformations $\delta^{(2)}\phi $ quadratic in fields) 
 which  are not known at present.
 While we will  not be able to compute the tadpole diagram,  the bubble diagram  already  provides    an  
 important information about the   self-energy  contribution  as  it captures  a non-trivial   part of its dependence on external momenta.
 Indeed,   the tadpole   contribution  is expected 
 to be a regular  function of external momentum   while  the bubble one 
should  contain branch cuts coming from logarithmic  terms appearing from 
massless  loops.



\subsection{Individual   bubble diagrams}

Let us start with the bubble  diagram with  two physical field   propagators in the loop. The momenta of particles ingoing the left vertex will be denoted 
 $p_1$  for the  external  scalar   and $p_2$ and $p_3$ for the 
spin $s_2$ and spin $s_3$ fields. 
Similarly, the momenta ingoing the
right vertex will be  $p'_1$  for the external scalar and $p'_2$ and $p'_3$ for higher-spin fields, i.e. 
\begin{align}
p_1+p_2+p_3 =0, \qquad p'_1+p'_2+p'_3=0,   \qquad 
p_1=-p'_1, \qquad p_2=-p'_2, \qquad p_3=-p'_3.  
\end{align}
 We shall also use the notation $k\equiv p_1$   for the external momentum and $p\equiv -p_2$  for the   virtual momentum. 
The  contribution of such bubble diagram reads\foot{We omit  the standard overall factor $(2\pi)^{-d}$ that
 should be reinstated in   the final    expression for the self-energy $\Sigma(k^2)$.}
\begin{eqnarray}
\notag
\begin{aligned}
\begin{fmffile}{fselfenergysmall}
\begin{fmfgraph}(20,15)
\fmfleft{i}
\fmfright{o}
\fmf{plain}{i,v1}
\fmf{photon,left,tension=0.3}{v1,v2,v1}
\fmf{plain}{v2,o}
\fmfdot{v1}
\fmfdot{v2}
\end{fmfgraph}
\end{fmffile}
\end{aligned} \quad = \quad  \ha \int d^dp_2\,  {\cal V}(\partial_{u_2},\partial_{u_3};p_1,p_2,p_3) \, 
{\cal V}(\partial_{u'_2},\partial_{u'_3};p'_1,p'_2,p'_3) \qquad\qquad
\\ \, \times\, {\cal D}_{s_2}^d(u_2,u'_2;p_2) \, 
{\cal D}_{s_3}^d(u_3,u'_3;p_3),
\label{looppic1}
\end{eqnarray}
where $\ha$ is a symmetry factor  and $s_2,s_3$   may  take any values including 0. 
  Here  ${\cal V}$   is the cubic vertex from \rf{8},\rf{416} 
and ${\cal D}_{s}^d$ is the propagator from \rf{Frprop1} (cf.  \rf{24}). 
The  dependence on $u$-variables  is spurious -- it goes away 
after acting by derivatives $\del_u$. For example,  
the left vertex ${\cal V}$ contains an operator
$(-i p_{31} \cdot \partial_{u_2})^{s_2}$
 which should be applied to  the propagator
${\cal D}_{s_2}$ thus performing the tensor  index contraction. 
Given that  
 ${\cal D}_{s_2}$ is a homogeneous polynomial 
of degree $s_2$ in $u_2$  (cf. \rf{20},\rf{010}) 
 the result of this contraction amounts to replacing
$u_2 \to -i p_{31}$ inside the propagator and also
 cancelling the factor of $s_2!$ in the denominator. 
Computing other  contractions in the same way we find 
 \begin{eqnarray}
 \begin{aligned}
\begin{fmffile}{fselfenergysmall}
\begin{fmfgraph}(20,15)
\fmfkeep{fselfenergysmall}
\fmfleft{i}
\fmfright{o}
\fmf{plain}{i,v1}
\fmf{photon,left,tension=0.3}{v1,v2,v1}
\fmf{plain}{v2,o}
\fmfdot{v1}
\fmfdot{v2}
\end{fmfgraph}
\end{fmffile}
\end{aligned} 
\quad = \quad
2 g_{0s_2s_3}^2 \int \frac{d^dp_2}{p_2^2 p_3^2}\,\frac{s_2!\,
C_{s_2}^{\frac{d}{2}-2}(1)}{2^{s_2} \left(\frac{d}{2}-2\right)_{s_2}}\,
\frac{s_3!\,C_{s_2}^{\frac{d}{2}-2}(1)}{2^{s_3} \left(\frac{d}{2}-2\right)_{s_3}}(p_{31}^2)^{s_2}(p_{12}^2)^{s_3} \ , 
 \end{eqnarray}
where $g_{0s_2s_3}$ is the cubic coupling constant  (to be chosen as in \rf{29})
 and  $C_{s_2}^{\frac{d}{2}-2}(1)$   is found  from   \rf{010},  i.e. 
\begin{equation}
C_s^{\alpha}(1) = \frac{\Gamma (2\alpha+s)}{\Gamma(2\alpha)\Gamma(s+1)} \ . \la{6115}
\end{equation}
We thus  obtain\foot{Note that the coefficient here is regular at $d=4$:
$
\frac{(d-4)_{s}}{ \left(\frac{d}{2}-2\right)_s} = \frac{(d-4) \;(d-3)_{s-1}}{\left(\frac d2-2\right)
\left(\frac d2-1\right)_{s-1}}\ \to \;2\ .  $}
 \begin{eqnarray}
 \begin{aligned}
 \fmfreuse{fselfenergysmall}
\end{aligned} 
\quad = \quad
2 g_{0s_2s_3}^2\,\frac{(d-4)_{s_2}}{2^{s_2} \left(\frac{d}{2}-2\right)_{s_2}}\,
\frac{(d-4)_{s_3}}{2^{s_3} \left(\frac{d}{2}-2\right)_{s_3}} \int \frac{d^dp_2}{p_2^2\,  p_3^2}\, (p_{31}^2)^{s_2}\, (p_{12}^2)^{s_3}\ .
\la{6.2} \end{eqnarray}
The ghost loop   contribution can be  computed in a similar way using \rf{24?},\rf{417},\rf{24}
\begin{eqnarray}
\notag
\begin{aligned}
\begin{fmffile}{gselfenergysmall}
\begin{fmfgraph}(20,15)
\fmfleft{i}
\fmfright{o}
\fmf{plain}{i,v1}
\fmf{ghost,left,tension=0.3}{v1,v2,v1}
\fmf{plain}{v2,o}
\fmfdot{v1}
\fmfdot{v2}
\end{fmfgraph}
\end{fmffile}
\end{aligned}
\quad=\quad
  g^2_{0s_2s_3}
 \,\frac{s_2 \,(d-2)_{s_2-1}}{2^{s_2-1} \left(\frac{d}{2}-1\right)_{s_2-1}}\,
\frac{s_3\,(d-2)_{s_3-1}}{2^{s_3-1} \left(\frac{d}{2}-1\right)_{s_3-1}}  \qquad\qquad 
  \qquad
\\
 \label{27} \times  \int \frac{d^dp_2}{p_2^2\,  p_3^2}\, 
{ (p_{31}^2)^{s_2-1}\,  (p_{12}^2)^{s_3-1}} (p_{31}\cdot p_2)\,  (p_3\cdot p_{12})\ .
\end{eqnarray}
Combining \rf{6.2} and \rf{27}   and  expressing the  momenta
in terms of   $k\equiv p_1$ and  $p\equiv -p_2$ we    get for the  bubble diagram    contribution with the 
spin  $s_2,s_3$  loop 
\begin{align}
\notag
\Sigma_{s_2 s_3} (k^2)= & g^2_{0s_2s_3} \int \frac{d^dp} {p^2 (p-k)^2}
 (2k-p)^{2s_2} (k+p)^{2s_3}\Big[2\,\frac{(d-4)_{s_2}}{2^{s_2} \left(\frac{d}{2}-2\right)_{s_2}}\,
\frac{(d-4)_{s_3}}{2^{s_3} \left(\frac{d}{2}-2\right)_{s_3}} 
\\
\label{28}
&\qquad\qquad+ \frac{s_2 \,(d-2)_{s_2-1}}{2^{s_2-1} \left(\frac{d}{2}-1\right)_{s_2-1}}\,
\frac{s_3\,(d-2)_{s_3-1}}{2^{s_3-1} \left(\frac{d}{2}-1\right)_{s_3-1}} 
\frac{ (2kp-p^2) (p^2-k^2)}{(2k-p)^{2} (k+p)^{2} }\Big]\ .
\end{align}

\subsection{Summing over spins}

Using the expression for  the coupling constants in  (\ref{29}) and  specifying to  $d=4$
we find  for   the sum $\Sigma  (k^2)$  of  the individual contributions   (\ref{28})  
\begin{align}
\notag
\Sigma (k^2)=  g^2 \int \frac{d^4p}{ p^2 (p-k)^2}
&\sum_{\substack{{s_2,s_3=0}\\{s_2+s_3={\rm even}}}}^\infty
\frac{\ell^{2(s_2+s_3-1)}}
{2^{s_2-1}2^{s_3-1} \big[(s_2+s_3-1)!\big]^2}\Big[2 (2k-p)^{2s_2}  (k+p)^{2s_3}
\\ 
\label{30}
&+s_2^2s_3^2
(2k-p)^{2(s_2-1)}  (k+p)^{2(s_3-1)}(2kp-p^2) (p^2-k^2)\Big] \ . 
\end{align}
Here  the sum goes over  even $s_2+s_3$   because  the vertex (\ref{8}) vanishes 
when the total spin  $s_2+s_3$  is odd  and the same is   also true for the
 ghost vertex \rf{18}.\foot{The cubic  vertex $S[\phi_0,\phi_{s_2},\phi_{s_3}] $  leads to the  two terms in the
ghost action,  schematically $V_1=S[ \bar c_{s_3-1}, c_{s_2-1},\p_0] $  and   $V_2= S[ \bar c_{s_2-1}, c_{s_3-1},\p_0]$
(see  \rf{15new}). 
Similarly, the   vertex $ S[\phi_0,\phi_{s_3},\phi_{s_2}] $ leads to 
$V_3=S[\bar c_{s_2-1}, c_{s_3-1},\p_0]$   and $V_4=S[ \bar c_{s_3-1}, c_{s_2-1},\p_0]$. 
One can check that $V_1 = (-1)^{s_2+s_3} V_4$ and $V_2 = (-1)^{s_2+s_3} V_3$, so that  for 
$s_2+s_3$=odd  the ghost action  vanishes.}
Evaluating the sum one finds (see Appendix C  for details)
\begin{align}
\la{338}
&\Sigma (k^2)=  g^2 \ll^{-2}  \int \frac{d^4p}{p^2 (p-k)^2} \   \Phi (p,k) \ ,  \\
&  \Phi(p,k)=
\Big(  \frac{4x^2 }{x-y}     +   \frac{2(2kp-p^2) (p^2-k^2)}{(2k-p)^2 (k+p)^2} \frac{4x^2 y^2 (x^2 + 4xy + y^2)}{(x-y)^5}  \Big)   
  \big[I_0(2\sqrt{x})-J_0(2\sqrt{x})\big]\no 
\\
\label{6.9}
&\qquad\qquad\quad   + \frac{2(2kp-p^2) (p^2-k^2)}{(2k-p)^2 (k+p)^2}
\Big(\frac{x^3 y (x+y)}{(x-y)^3}\big[ I_2 (2\sqrt{x})-J_2 (2\sqrt{x})\big] \\
&\qquad\qquad\qquad\qquad\qquad+\frac{2x^2y(x^2 - 8xy-5y^2)\sqrt{x}}{(x-y)^4}   
\big[I_1(2\sqrt{x})+J_1(2\sqrt{x})\big] \Big) 
+  ( x \leftrightarrow y )\ ,   
\no 
\end{align}
where  $I_n$ and $J_n$ are  the Bessel functions \rf{5.7}   and we defined 
\begin{align}\la{610}
x\equiv  \ha  {\ell^2 (2k-p)^2}, \qquad & \quad y \equiv \ha  {\ell^2(k+p)^2}\ . 
\end{align}

\subsection{UV divergences and  regularization}
Let us   now discuss the UV  divergence  of $\Sigma$. 
There are at least  two possible approaches:  (i) first introduce   an explicit momentum 
 UV cutoff,   sum over spins  and then  take cutoff to infinity,  
or (ii)  first  drop all power divergences in each  individual  loop  using the dimensional regularization and then 
 sum over spins.\foot{In general, dimensional regularization 
 is known to be a   preferable choice in order  to preserve  gauge invariance of the theory at the quantum level 
 but in an effectively  non-local theory like the present one  its use  needs to be further justified, e.g., it 
  may not commute with  summation over spins (see also below).}

Following the first approach,  to isolate  the UV   divergence  of  the loop integral in  \rf{6.9}   let us consider 
the   $p\to \infty, \ k$=fixed   limit   of the integrand. 
In this limit  ${x\ov y}\to 1$.  To extract the leading singularity one may just set $k=0$. 
 The  apparent   $x=y$   poles of the   integrand are  spurious  and one finds (see Appendix C) 
  \begin{align}
\notag
 \Sigma (0)=&\; 2 g^2 \ll^{-2}  \int \frac{d^4p}{p^4}\ 
 \te \Big(-\frac{1}{30}\big(\frac{\ell p}{{ \sqrt{2}}}\big)^7 \big[I_5(\sqrt{2}\ell p)+J_5(\sqrt{2}\ell p)\big]
-\ha\big(\frac{\ell p}{{ {\sqrt{2}}}}\big)^6 \big[I_4(\sqrt{2}\ell p)-J_4(\sqrt{2}\ell p)\big]
\\
\notag
&\qquad \te -\frac{13}{6}\big(\frac{\ell p}{{ \sqrt{2}}}\big)^5 \big[I_3(\sqrt{2}\ell p)+J_3(\sqrt{2}\ell p)\big]
-3\big(\frac{\ell p}{{ \sqrt{2}}}\big)^4 \big[I_2(\sqrt{2}\ell p)-J_2(\sqrt{2}\ell p)\big]\\
&\qquad \te +\big(\frac{\ell p}{{\sqrt{2}}}\big)^3 \big[I_1(\sqrt{2}\ell p)+J_1(\sqrt{2}\ell p)\big]+
4\big(\frac{\ell p}{{ \sqrt{2}}}\big)^2 \big[I_0(\sqrt{2}\ell p)-J_0(\sqrt{2}\ell p)\big]   \Big) \ .
 \label{616}
 \end{align}
From  the asymptotic     behaviour of the Bessel functions \rf{588} 
we  conclude  that  $\Sigma$ in  (\ref{338}) is exponentially  UV divergent, 
i.e.   introducing an  explicit UV momentum cutoff $\Lambda$  
we get 
\be \la{611}
 \Sigma (0)\sim g^2 \ell^{-2 }\int^\Lambda{  d^4p\ov p^4 }  \ \Big( \big[ (\ll p)^{13/2}   + ...\big]  \, e^{\sqrt 2 \ll p } + ... \Big) \sim 
g^2  \ell^{-2 } (\ll \Lambda)^{11/2} \, e^{\sqrt 2 \ll \Lambda} + ... \ . 
 \ee
Keeping also the subleading  order $k^2$ term  in the large $p$ expansion of the integrand in \rf{338}  gives 
\begin{align}
 &\Sigma(k^2) =
  g^2  \ell^{-2 } \int^\Lambda  \frac{d^4p}{p^4}\te \Big( \Big[ -\frac{1}{15}(\frac{\ell p}{\sqrt{2}})^7+
 \frac{1}{15}(\frac{\ell p}{\sqrt{2}})^8\frac{kp}{p^2} -\frac{8}{105}(\frac{\ell p}{\sqrt{2}})^9(\frac{kp}{p^2})^2 
 \no \\&
 \te 
 -\frac{1}{6}(\frac{\ell p}{\sqrt{2}})^8\frac{k^2}{p^2} + ...\Big]  \frac{   e^{\sqrt{2}{\ell p}}   }{2^{3/4}\sqrt{\pi{\ell p}}} + ... \Big)  
 \sim
 {g^2 \ell^{-2 }}\big(1+\frac{1}{7} \ell^2 k^2 +  ...\big) \left({\ell \Lambda}\right)^{11/2}
 {   e^{\sqrt{2}{\ell \Lambda}}   } + ... \ . \la{614}
 \end{align}
 Note also that the logarithmic $\log {\Lambda \ov k} $ divergence in \rf{338},\rf{614}   has a  non-zero  coefficient 
 \be \la{4112} 
 \Phi(0,k)=
 \te  \frac{2}{3} \ell^2 k^2\Big(     16  \big[I_0(2\sqrt{2}\ell k )-J_0(2\sqrt{2}\ell k )\big]     -        \big[I_0(
  \sqrt{2}\ell k )-J_0(\sqrt{2}\ell k )\big]   
\Big) \ . 
  \ee
 

Instead of using the UV  momentum cutoff  in  the summed over spins expression 
one may  first  define each loop  integral \rf{28} 
 with the help of   dimensional regularization  which gets rid of all power divergences, i.e.  sets 
\begin{equation}
\label{dr1}
\int {d^4p}\left(p^2\right)^n=0, \qquad\qquad  n=-1,0,1,... \ . 
\end{equation}
Then for  integer  $n=-1,0,1...$  and $m=0,1, ...$ (or vice versa) one has also  
\begin{equation}
\label{dr11}
 \int {d^4p}\left(p^2\right)^m \left[(p-k)^2\right]^n  = 0 \  . 
\end{equation}
\iffa 
Since the power $n$ is not negative the integrand is polynomial in $k_a$.
One can argue by symmetry that 
odd powers of $k$ drop out upon integration, while  $k$'s in $k$-even terms get contracted with the
metric. 
Schematically,
\begin{equation*}
k_{a_1} k_{a_2} \dots k_{a_{2j}}\qquad  \to \qquad \left(k^2\right)^j \eta_{a_1a_2} \dots \eta_{a_{2j-1}a_{2j}},
\end{equation*} 
where we
omitted an unimportant 
numerical prefactor.
Any $k^2$ dependence can be pulled off the integration sign, while each metric tensor being 
contracted with $p^a$ produces 
$p^2$. Thus,   (\ref{dr11}) reduces to a certain combination of integrals of the form (\ref{dr1}) 
and hence vanishes
\begin{equation}
\label{dr2}
{\cal I}_1(m,n)=0, \qquad n\ge 0.
\end{equation}
 Analogously, shifting the integration variable $p\to (p-k)$ one can show that
\begin{equation}
\label{dr3}
{\cal I}_1(m,n)=0, \qquad m\ge 0.
\end{equation}
\fi
The integral (\ref{30}) has the  general  form  
\begin{align}
\label{dr4}
\Sigma(k^2) = g^2   \ell^{-2 } \int \frac{d^4p}{p^2(p-k)^2}\,  \Phi \big(p^2,(p-k)^2,k^2 \big) \ , 
\end{align}
where the function $\Phi  $  is a series in  positive integer powers of its 
arguments. Then  assuming \rf{dr11}  we  conclude  that 
only $ \Phi(0,0,k^2)\equiv \Phi(k^2)$ produces a non-zero  contribution 
in dimensional regularisation, i.e. we   are left with only  logarithmically divergent integral   
\begin{equation}
\label{dr5}
\Sigma(k^2) =g^2 \ll^{-2} \,  \Phi(k^2) \int \frac{d^4p}{p^2 (k-p)^2} \ , \qquad  
\Phi(k^2) = \sum_{\substack{{s_2,s_3=0}\\{s_2+s_3={\rm even}}}}^\infty
\frac{(8-\, s_2^2 \, s_3^2)\, (\ell^2 k^2)^{s_2+s_3}\, }{\big[(s_2+s_3-1)!\big]^2} \ . 
\end{equation}
Computing the sum  as discussed in Appendix C 
we get  \begin{align}
\notag
 \Phi (k^2)=  &\te   -\frac{1}{60}\left({\ell k}\right)^7 \big[  I_5({2}\ell k)+J_5({2}\ell k)\big]
-  {1  \ov 4} \left({\ell k}\right)^6 \big[ I_4({2}\ell k)-J_4({2}\ell k)\big] 
\\
\label{632}
&\te -\frac{13}{12}\left({\ell k}\right)^5 \big[I_3({2}\ell k)+J_3({2}\ell k)\big]
-{3\ov 2} \left({\ell k}\right)^4 \big[I_2({2}\ell k)-J_2({2}\ell k)\big]
\\&\te +{7\ov2}\left({\ell k}\right)^3 \big[I_1({2}\ell k)+J_1({2}\ell k)\big]+
8\left({\ell k}\right)^2 \big[ I_0({2}\ell k)-J_0({2}\ell k)\big] \ .   
\notag
 \end{align}
 Thus with this  dimensional regularization  prescription  (where one discards all power divergences 
    before summation over spins) 
 one finds  that instead of  an  exponential divergence  in \rf{611}  
  the self-energy  divergerges only logarithmically\foot{
  The  self-energy diagram   we are computing is an off-shell  one, and 
  as in scalar electrodynamics this  logarithmic divergence  may be  absorbed into
  a  wave  function renormalization of the spin 0  field. 
  }
     and, in particular, 
    vanishes  at zero momentum,  $\Sigma(0)=0$, so that 
   the spin 0 field remains massless.

 
 \subsection{Comments on   tadpole diagram  contribution   }
 
 One may wonder if the exponential divergence \rf{611}   may get cancelled   upon adding the tadpole 
 diagram contribution $  \Sigma_{\rm tp} (k^2) $. As we have seen  in section 5,  the expected structure of the 4-vertex \rf{59} 
  leads to a contribution  to the 4-scalar tree-level scattering amplitude that  has  
  similar exponential   large momentum   behaviour \rf{62} as the exchange  diagram \rf{58}. 
 One may thus  expect   that this 4-vertex contribution to the self-energy tadpole   diagram  will also 
 have an exponential UV   behaviour similar to the one in \rf{611}. 
 
 Let us consider just a  single   virtual $s=0$  field   contribution to  the tadpole   diagram 
 for which the knowledge of the 0-0-0-0  vertex \rf{59} is sufficient.   
  Computing the scalar loop   with two scalar  external legs we then get  
  the following   estimate for its  large virtual momentum    behaviour   (using \rf{60},\rf{61}, cf. \rf{62})
 \begin{equation}\la{64}
\Sigma_{\rm tp} (k^2)\sim  g^2  \int \frac{d^4 p }{p^2} \sum_{j=0} \Big[f_{2j}\big(- (k-p)^2\big) \big[(k+p)^2\big]^{2j} + 
f_{2j} \big(-(k+p)^2\big) \big[(k-p)^{2}\big]^{2j}\Big] \ .
\end{equation}
Setting $k=0$ to get  the leading UV  asymptotics   we   find  
 \begin{equation}\la{65}
 \Sigma_{\rm tp} (0) \sim  g^2   \int \frac{d^4 p}{p^2} \sum_{j=0}a_{2j}(-p^2) \left(p^{2} \right)^{2j}
 = \fo   g^2  
 \int \frac{d^4 p}{p^2}\,  \Big[ I_0(\sqrt{2}\ell p)-J_0(\sqrt{2} \ell p)\Big] \ , 
 \end{equation}
 where $a_{2j}(z) ={ \ell^{4k} \ov  2^{2k} [(2k-1)]^2}{ 1 \ov z} $.
 Thus  this  tadpole diagram   has  a   similar  exponential UV behaviour
 as the  bubble diagram \rf{6.9}. 
 
 This gives a  hope   that the UV  divergence  of the bubble  diagram   contribution   may get cancelled
 once all tadpole diagrams  (for all spins propagating in the loop) are taken into account. 
 Ideally, the sum of  the bubble and tadpole contributions  may  be given   
 by a convergent momentum integral that will not require UV regularization 
   and may actually vanish.\foot{The prescription where  
    one first   combines  all contributions together,  sums over spins  
   and only then  discusses the need for a UV  regularization  seems the most natural one. 
   The  application of dimensional regularization to individual graphs  
   may be objected as it  may not  commute with 
    summation over spins. For example, 
   a convergent integral  like 
   $\int  { d^4p \ e^{-\ll^2p^2} \ov p^2(k-p)^2} $  may be   represented as a  sum of divergent integrals 
   $\sum_{n=0}^\infty {(-1)^n\ov n!}  \int  { d^4p \   {(\ll^2 p^2)}^n \ov p^2(k-p)^2} $ 
   with  all $n\geq 1$ terms vanishing if computed   in dimensional regularization  but  the remaining $n=0$ one having 
   the logarithmic  divergence.}
   The same  may be true also for the case of  self-energy diagram with an arbitrary spin  $s$ field on external lines. 
 Such a result is expected in the AdS  higher-spin theory dual to a  free boundary  $U(N)$ or $O(N)$  scalar   CFT
 where the   one-loop self-energy correction would represent a   $1/N$ correction to the  2-point function of  conserved currents
 which should be absent  in a free theory (i.e. a  theory with unbroken higher spin   symmetry). 
 

   
\section{Concluding remarks}

In  this paper  we  computed the one-loop bubble diagram correction to the scalar propagator 
generated by interactions with an infinite tower of higher-spin gauge fields in flat 4d space.
One  motivation was to investigate whether in higher-spin theories, similarly to what happens in 
string theory, the summation over an infinite number of spins may 
 make loop integrals finite in the ultraviolet. 
  Another is that this     may   be considered 
 as a simplified version of the analogous computation in the massless AdS higher-spin theory,
 aiming to verify a remarkable prediction of the  vectorial AdS/CFT duality  that all loop corrections
 in this  higher-spin theory should vanish. 
 
 The explicit cubic coupling coefficients \rf{29}  that we used were previously
 derived by demanding consistency of higher-spin interactions to the  subleading $g^2$ order 
  in  the light-cone approach  \cite{Metsaev:1991mt}. 
  We also used these coefficients  to compute the tree  scattering 
   amplitude of the massless scalars   due  to the exchange   of   an infinite  tower of  massless  higher spins.
  We found that this exchange amplitude has an exponential  growth in the 
   large momentum  limit, 
  suggesting   singular UV behaviour in the loops. Indeed, the bubble diagram  contribution to the scalar self-energy 
 (summed over all virtual spins propagating in the  loop)
   was found to be  exponentially divergent at high energies. 
 
Qualitatively, this happens because all contributions from different spins (each of which grows in the UV due to 
the presence of higher derivatives in the cubic vertex) enter 
with the same sign and  thus   the summation over spins cannot improve
the ultraviolet behaviour.  The  external  spin 0  field  we  were scattering 
 is  a member of the higher-spin tower, i.e. a  real scalar   which couples   via 0-0-$s$  
  vertex only to even spin $s$ fields.\foot{At the same   time,  if one  considers  the  scattering  of a complex  scalar that couples also  to odd spins (with extra $i$ in the vertex) 
one may get    softer UV  behaviour  due to alternating signs of coefficients  in  sum over all spins  
 \cite{Bekaert:2009ud}. 
 It is not clear,  however,  if there is a consistent  higher-spin theory (with complex scalars   being  part 
 of the spectrum)   where this  UV softening   mechanism  can be implemented  also  at the  loop level. In particular, 
 it is not clear if  similar  alternating series  may appear in the bubble   diagrams depending on cubic  vertices with two  higher spin fields.}  
 

In addition to  the exchange  diagrams,   the full tree-level 4-scalar amplitude should   contain
 also the contribution of the 4-point vertex. 
Similarly,   besides the 
 bubble diagrams, the full  one-loop self-energy  correction  should contain
  also the contributions of the  tadpole
diagrams. To compute these  extra contributions  requires  the   knowledge  of 
quartic vertices  which are not   understood at present. 
To get an idea  of   possible  structure of 4-vertex  contributions   we used  the 4-scalar  interaction term 
found in the AdS higher-spin theory   \cite{Bekaert:2015tva} and formally continued it  to flat space in the short distance limit. 
 We found   that it  leads, indeed,  to   similar UV  behaviour as the tree-level exchange diagrams 
 and also to  similar  tadpole   momentum  integrals as   appear in the bubble diagram. 
  There is  thus  a potential  possibility  of cancellation  of   UV  divergences in
  the full one-loop self-energy correction.  
  This   is    what  is  to   happen in the AdS  higher-spin theory dual to a  boundary CFT 
    and  would be in line 
  with the  expectation that the UV behaviour  in  the flat space and AdS   theories should be similar. 
 

\iffa 
Finally, let us make a suggestive observation. Assuming that there is a dictionary between
processes in the bulk of AdS and processes in flat space as the one established in \cite{Penedones:2010ue},
one expects that the AdS/CFT prediction of cancellation  of bulk loop Witten diagrams
translates into the requirement that 
on-shell loop Feynman diagrams also cancel in flat space.
 Dropping terms that vanish in dimensional regularisation,
we find that the scalar self-energy that we evaluated is indeed zero on-shell. Computationally,
this  relies on the fact that  Metsaev's cubic scalar self coupling vanishes. In AdS${}_4$
the cubic 
scalar self coupling is also zero, as this is required by matching with the  3-point correlator of the CFT.
 However, it is no longer the case in other dimensions.
\fi

\iffa 
*******
Future problems:

one-loop box  diagram for 4 scalars  (at one loop other connected diagrams are  not 1-PI) 

l.c. gauge computation -- known vertices of Metsaev 

symmetry constraints

BCFW    relation 
\fi 

\section*{Acknowledgments}

We thank X. Bekaert, S. Giombi and  E. Joung for discussions and are also   grateful  to 
 R. Metsaev, R. Roiban and E. Skvortsov for  very useful  comments on the draft.
 This work
was  supported by the ERC Advanced grant No.290456. The work of AAT was also supported 
by the STFC Consolidated grant  ST/L00044X/1 and the 
RNF grant 14-42-00047.

\def \sq {{\te {1\ov \sqrt 2}}}
\def \x {{\rm x}} \def \bx {\bar {\rm x}}
\def \third {{\te {1\ov 3}}}
\def \ev {{\rm even}}


\

\appendix
\section{Relation  between covariant and light-cone cubic  vertices}

Our aim here will be to establish the   relation   between the covariant cubic vertex \rf{8},\rf{29}   and the light-cone gauge one 
in   \cite{Metsaev:1991mt}.
We shall consider the case of $d=4$ Minkowski space   with metric $\eta_{ab} = \rm{diag}(-,+,+,+)$
and define 
\begin{align}
&x^{\pm}= \sq (x^0\pm  x^3) \ ,\qquad\qquad\qquad   \x= \sq (x^1+ i x^2)\ , \qquad \bar \x= \sq (x^1- i x^2) \ , \no \\
& \partial^{\pm}=-\del_{\mp} = \sq (-\partial_0\pm \partial_3)\ , \qquad 
\partial=\sq (\partial_1 + i  \partial_2)\ , \qquad 
 \bar\partial=\sq (\partial_1 - i  \partial_2)\ ,   \la{a1}
\end{align}
so that  $ds^2 = -2dx^+ dx^-  +2 d\x d\bar \x.$
Below  we will label the 4d  indices by $a=0,1,2,3$ and  by $(+,-,\x,\bar \x)$ and also   use $I,J,...$ to label the 
$\x$ and $\bar \x$  directions. 
We will  follow  the standard notation
\begin{equation}\la{a2}
\phi^{a(s)} \equiv \phi^{a_1 a_2\dots a_s} \ , \qquad \qquad 
\big(\partial_{x}\big)_{a(s)} \equiv \frac{\partial}{\partial x^{a_1}}\frac{\partial}{\partial x^{a_2}} \dots
\frac{\partial}{\partial x^{a_s}}.
\end{equation}
that is instead of writing all indices of the symmetric tensor, we just write one of them and indicate
their number in brackets.

\subsection{Free fields in light-cone gauge}

 The  light-cone gauge  for the Fronsdal fields that fixes the gauge freedom 
 (\ref{3}) completely  corresponds to  setting  to zero all the components of the off-shell 
field $\p^{a(s)}$ that  have  at least one upper ``$+$" index  
\begin{equation}
\label{lc1}
\phi^{+ -(n) I (s-n-1)} = 0  \ .  
\end{equation}
Here $n$ is the number of ``$-$" indices and  the remaining $s-n-1$ indices  are the transverse ones $I=(\x,\bx)$.
In the light-cone gauge approach one usually assumes that derivatives $\partial^+$ of all
fields and gauge parameters are non-vanishing, i.e.  one can divide by $\partial^+$.\foot{Inverse  powers of 
 $\partial^+$ need not be  considered as an  indication of  genuine non-localities 
 but  originate from solution of on-shell constraints.}

Let us review the  consequences  of this gauge condition  when combined with 
partial set  of the   equations of motion (\ref{eom}) that allow one  to  express  all of the components 
of the Fronsdal  field in terms of just two independent ones. 
The components  of e.o.m.  (\ref{eom})   with two ``$+$"  indices give 
\begin{equation}
\label{lc2}
\partial^+\partial^+ \phi_m{}^{ma(s-2)}=0 \quad \Rightarrow \quad  \phi_m{}^{ma(s-2)}=0 \ , 
\quad {\rm i.e. } \quad  
\phi_I{}^{Ia(s-2)}=0 \ . 
\end{equation}
This  implies that there are  only two non-vanishing components of $\phi^{I(s)}$ 
\begin{equation}
\label{lc3}
\varphi_s\equiv \phi^{\x(s)} \ , \qquad \qquad \bar\varphi_s\equiv \phi^{\bar \x(s)} \ .
\end{equation}
The components  of   (\ref{eom})   with one  ``$+$"  index  give 
\begin{equation}
\label{lc4}
\partial^+ \partial_m \phi^{m a(s-1)}=0 \quad \Rightarrow \quad \partial_m \phi^{m a(s-1)}=0\ ,
\end{equation}
so the field is also transverse. Special cases of  \rf{lc4} give 
\begin{equation}
\phi^{-I(s-1)} = \frac{\partial_J}{\partial^+}
 \phi^{JI(s-1)} \ , 
%
 \quad \phi^{--I(s-2)} = \frac{\partial_J}{\partial^+}
 \phi^{-JI(s-2)} \ , \quad 
\label{lc6}
 \phi^{--I(s-2)} = \frac{\partial_J \partial_J}{(\partial^+)^2}
  \phi^{JJI(s-2)}.
\end{equation}
Proceeding in the same manner one gets
\begin{equation}
\label{lc7}
 \phi^{-(k)I(s-k)} = \frac{1}{(\partial^+)^k}
(\partial_J)^k  \phi^{J(k)I(s-k)}.
\end{equation}
Thus 
there are  only two on-shell independent components of
the $d=4$  Fronsdal  field in  the light-cone gauge  --   the two helicity fields (\ref{lc3}). The equations of motion for them 
have the standard $\del^a\del_a =\Box $  kinetic term, i.e.  follow from the action\foot{It is interesting to note 
the light-cone gauge action for free Fronsdal fields in AdS$_4$   also has the same form as \rf{lc8} 
\ci{Metsaev:1999ui}, i.e. there are no extra ``mass" terms present in the covariant expression (such terms appear though 
  in $d >4$).}
\begin{equation}
\label{lc8}
S_2[\varphi_s,\bar\varphi_s]= \int d^4x\; \bar\varphi_s \Box \varphi_s .
\end{equation} 

\subsection{Cubic interactions}

One  imposes the same  light-cone gauge  (\ref{lc1}) also at the interacting level.
Interactions deform  the free  equations in a way  that  (\ref{lc2}), (\ref{lc4}), (\ref{lc7})  now hold only 
 up to terms linear in the coupling
constant, e.g., 
\begin{equation}
\label{lc9}
\phi_m{}^{ma(s-2)}={\cal O}(g)\ ,
\qquad \qquad 
\partial_m \phi^{m a(s-1)}={\cal O}(g) \ .
\end{equation}
These equations should be again used to express all of the auxiliary  components of $\p_s$  in terms of the two dynamical 
helicity 
fields $\varphi_s$ and $\bar\varphi_s$.
The  elimination of the  auxiliary fields generates  higher powers of $\varphi_s$ and $\bar\varphi_s$ 
even from the quadratic action but they contribute only to  quartic  and  higher order 
 interactions.
Indeed, noting that except for  the $\partial_x^2$-term all other terms in the free action 
 (\ref{1b}) are at least bilinear in traces and divergences we  conclude  that
\begin{equation}
\label{lc10}
{\te \frac{s!}{2}}\int d^4 x\; \left[ \phi_s(x,\partial_u) \,\hat T\,\hat {\cal F} \phi_s(x,u)
\right]_{u=0} = 
 \int d^4x\; \bar\varphi_s \Box \varphi_s  +{\cal O}(g^2).
\end{equation}
Thus to  find the  cubic vertices for $ \varphi_s $ field 
 in the light-cone gauge from those in the covariant approach one just needs to plug (\ref{lc1})-(\ref{lc7})
 into the  covariant cubic action.

Let us start with the traceless-transverse part of the highest derivative
 cubic vertex  in 4 dimensions
\cite{Manvelyan:2010jr,Sagnotti:2010at,Manvelyan:2010je,Joung:2011ww}\foot{This vertex contains the maximal number of derivatives consistent with 
the condition of being non-zero when the  fields are restricted to  solutions of free equations  of motion.} 
\begin{align}
\notag
S^{(3)}[\phi_{s_1},\phi_{s_2},\phi_{s_3}] = g_{s_1s_2s_3}\int d^4 x & \Big[ 
(\partial_{u_1} \cdot \partial_{x_{23}})^{s_1}
(\partial_{u_2} \cdot \partial_{x_{31}})^{s_2}
(\partial_{u_3} \cdot \partial_{x_{12}})^{s_3}
\\
\label{lc11}
&\qquad  \times  \phi_{s_1}(x_1,u_1)\phi_{s_2}(x_2,u_2)\phi_{s_3}(x_3,u_3)\Big]_{\substack{{u_i=0}\\{x_i=x}}} \ .
\end{align}
The vertex in \rf{8} is the special case of \rf{lc11} corresponding to $s_1=0$. 
In the light-cone gauge one finds\foot{Here in the last line we assume the possibility of   integration   by parts, 
i.e. $\del_3 \to -\del_1 - \del_2$.}
\begin{align}
&\big(\partial_{x_{12}}\big)_{a(s_3)} \phi^{a(s_3)}(x_3)=
\sum_{n=0}^s \frac{s!}{n!(s-n)!}\,\big(\partial_{x_{12}}\big)_{-(n)}\big(\partial_{x_{12}}\big)_{I(s_3-n)}\phi^{-(n)I(s_3-n)}
\no \\
&=\sum_{n=0}^s \frac{(-1)^ns!}{n!(s-n)!}\,\big(\partial_{x_{12}}\big)^{+(n)}\big(\partial_{x_{12}}\big)_{I(s_3-n)}
\frac{\partial_{3 J(n)}}{(\partial_3^+)^n}\,\phi^{J(n)I(s_3-n)} \no 
\\
\no
& =\sum_{n=0}^s \frac{(-1)^ns!}{n!(s-n)!}\,(\partial^+_{x_{12}})^{n}\bar\partial_{12}^{s_3-n}
\frac{\bar\partial^n_3}{(\partial_3^+)^n}\,\varphi_{s_3}
 + \sum_{n=0}^s \frac{(-1)^ns!}{n!(s-n)!}\,(\partial^+_{x_{12}})^{n}\partial_{12}^{s_3-n}
\frac{\partial^n_3}{(\partial_3^+)^n}\,\bar\varphi_{s_3}
\\
& =\Big(\bar\partial_{12}- \frac{\partial^+_{12}\bar\partial_3}{\partial_3^+}\Big)^{s_3}\varphi_{s_3}+
\Big(\partial_{12}- \frac{\partial^+_{12}\partial_3}{\partial_3^+}\Big)^{s_3}\bar\varphi_{s_3} \no \\
 \label{lc12} 
& =
\Big(2 \frac{\bar \partial_2 \partial_1^+ - \bar \partial_1 \partial_2^+}{\partial_3^+}\Big)^{s_3}\varphi_{s_3}+
\Big(2\frac{ \partial_2 \partial_1^+ -  \partial_1 \partial_2^+}{\partial_3^+}\Big)^{s_3}\bar\varphi_{s_3}.
\end{align}
Following 
 \cite{Metsaev:1991mt}  let us  introduce the notation 
\begin{equation}
\mathbb{P}\equiv \third \big[  {\partial_1(\beta_2-\beta_3)+\partial_2(\beta_3 - \beta_1)+\partial_3 (\beta_1-\beta_2)} \big] \ , 
\qquad \qquad \beta \equiv \partial_- \ .
\end{equation}
It is easy to see that
\begin{equation} \frac{\bar \partial_2 \partial_1^+ - \bar \partial_1 \partial_2^+}{\partial_3^+} = - \frac{\bar{\mathbb{P}}}{\beta_3} \ ,
\end{equation}
so that  (\ref{lc12})  can be rewritten as 
\begin{equation}
\label{lc13}
(\partial_{x_{12}})_{a(s_3)} \phi^{a(s_3)}(x_3)=\Big(-2 \frac{\bar{\mathbb{P}}}{\beta_3}\Big)^{s_3}\varphi_{s_3}+
\Big(-2 \frac{{\mathbb{P}}}{\beta_3}\Big)^{s_3}\bar\varphi_{s_3}\ .
\end{equation}
As a result, we find that   the covariant cubic vertex (\ref{lc11}) written in the light-cone gauge becomes
\begin{equation}\la{a17}
S_3 = g_{s_1s_2s_3}\int d^4 x\  \Big(-2 \frac{\bar{\mathbb{P}}}{\beta_1}\Big)^{s_1}\Big(-2 \frac{\bar{\mathbb{P}}}{\beta_2}\Big)^{s_2}\Big(-2 \frac{\bar{\mathbb{P}}}{\beta_3}\Big)^{s_3}\varphi_{s_1}\varphi_{s_2}\varphi_{s_3}+\dots,
\end{equation}
where dots stand  for  analogous terms involving $\bar\varphi$.
 The  light-cone gauge vertex found in   \cite{Metsaev:1991mt} has  exactly this  form with\foot{The 
  result (\ref{coup}) was derived in \cite{Metsaev:1991mt} for the special case when 
  all spins entering the vertex are even and the more 
   general case (with symmetric/antisymmetric internal indices on even/odd spin 
   fields$\p_s$) was considered in \cite{Metsaev:1991nb}. We shall assume that   \rf{a17}  holds 
for all values of the spins.}  
\begin{equation}
\label{coup}
g_{s_1s_2s_3} = g \frac{\ell^{s_1+s_2+s_3-1}}{(s_1+s_2+s_3-1)!} \ ,
\end{equation}
where $g$  is an overall  coupling constant  and $\ll$ is an arbitrary dimension-length parameter.

Let us note   that the vertex (\ref{lc11}) has the following
symmetry under the  exchange of any two arguments
\begin{equation}
S^{(3)}[\phi_{s_1},\phi_{s_2},\phi_{s_3}] = (-1)^{s_1+s_2+s_3}S^{(3)}[\phi_{s_2},\phi_{s_1},\phi_{s_3}]\ . 
\end{equation}
To get a complete cubic interaction,  this vertex  has to be summed over all fields  and this   implicitly
 symmetrizes  it  over its arguments. This implies that the
cubic  vertex (\ref{lc11})  contributes  under  summation over all spins only if the total spin 
of the fields is even, i.e. we may assume   that  the vertex  with the total spin being odd 
may be effectively set to zero 
\begin{equation}
\label{totalodd}
S^{(3)}[\phi_{s_1},\phi_{s_2},\phi_{s_3}] = 0 \qquad \text{if} \qquad s_1+s_2+s_3 = {\rm odd} \ . 
\end{equation}

\section
{Test of BCFW  constructibility condition  }

In this Appendix we  shall  discuss if the  above higher spin  cubic vertices 
can be  used to derive the quartic ones within  the BCFW framework, i.e. if the BCFW  constructibility condition is satisfied
(cf. \ci{Benincasa:2007xk,Benincasa:2011pg}). 

Let us consider a  formal   shift of two momenta 
\begin{equation}
\label{bcfw1}
p_i \to p_i(w)=p_i +w q \ , \  \ \qquad p_j \to p_j(w)=p_j -wq \ ,
\end{equation}
where   $w$  is a complex number   and the vector $q$ satisfies
\begin{equation}
\label{bcfw2}
q^2=0\ ,\ \qquad q\cdot p_i = 0 \ , \qquad q\cdot p_j=0\   .
\end{equation}
  The  BCFW constructibility condition \cite{Benincasa:2007xk} 
is  that 
under   such  shift   the total 4-point amplitude  should vanish in the limit $w\to \infty$, 
\begin{equation}
\label{bcfw3}
\lim_{w\to \infty} \hat {\cal A}(w)=0\  . 
\end{equation}
Then  assuming  $\hat {\cal A}(w)$ is analytic it   may have  only poles at   finite values of $w$.
 These  poles   should  correspond  to  the values of $w$ for which 
the internal propagators 
that involve shifted momenta go on-shell. 
The residues at these poles are given  by products of  on-shell amplitudes
resulting from the original one  after  cutting the propagator that goes on-shell under
the  shift. Using analyticity of  the shifted amplitude
${\cal A}(w)$ as a function of $w$, one can then  recursively express it in terms of the 
products of on-shell three-point amplitudes.

Let us  apply these   considerations  to  the  tree-level  four-scalar  amplitude   discussed  in section 5. 
Under the  shift \rf{bcfw1}   applied   to $p_1$ and $p_2$   in \rf{51}   we find  for the Mandelstam variables in \rf{511} 
\begin{equation}
\label{bcfw4}
 \s(w) =\s ,\qquad  \t(w)= \t+w \Delta  , \qquad \u(w) = \u-w\Delta ,\qquad \qquad \Delta\equiv -2 q\cdot p'_1\ . 
\end{equation}
Then the constructibility condition (\ref{bcfw3}) implies
that the total 4-scalar amplitude  containing the exchange and ``contact" (4-vertex) contributions 
\be\la{bb1}
  \hat {\cal A}(\s,\t,\u)=\hat {\cal A}_{exch}(\s,\t,\u) + {\cal A}_{cont}(\s,\t,\u)   \ , \ee 
should  vanish  at $\t\to\infty$.
 The total exchange
part  \rf{577} 
of the  4-point  amplitude   corresponding to  the $\s$-channel expression in  (\ref{57}),\rf{56}  is 
 \begin{align}
 \notag
\hat  {\cal A}_{exch}(\s,\t,\u)= 
 -&\frac{i}{\s}\Big[ F\Big(\ha{(\sqrt{-\u}+\sqrt{ \t}\, )^2}\Big) + F\Big(\ha{(\sqrt{-\u}-\sqrt{\t}\, )^2}\Big) 
  \Big] \\
  \notag
  -&\frac{i}{\t}\Big[ F\Big(\ha{(\sqrt{-\u}+\sqrt{ \s}\, )^2}\Big) + F\Big(\ha{(\sqrt{-\u}-\sqrt{\s}\, )^2}\Big) 
  \Big]\\
  \la{bcfw5}
  -&\frac{i}{\u}\Big[ F\Big(\ha{(\sqrt{-\s}+\sqrt{\t}\, )^2}\Big) + F\Big(\ha{(\sqrt{-\s}-\sqrt{\t}\, )^2}\Big) 
  \Big] \, . 
 \end{align}
To satisfy the constructibility condition, the  contact  contribution  should cancel the exponential 
 singularity   \rf{58} of this  exchange amplitude 
\rf{bcfw5}  at $\t\to \infty$. 
  Given that $F(z)$ in \rf{56}  is an entire function, then   (under a natural assumption  that ``contact" contribution should not  introduce new  poles)  the only remaining singularities  of  \rf{bcfw5}  will   be  
 poles   at $\t=0$ and $\t=-\s$.
Keeping only the contributions from these  finite poles (as required by the assumption of  BCFW constructibility), 
the total amplitude    would then be 
\begin{equation}
\label{bcfw55}
\hat {\cal A}'(\s,\t,\u)= -i   \frac{\s}{\t(\s+\t)}F(2\s)  \ ,
\end{equation}
where   we used that $F(z) = F(-z)$ (see \rf{56}).
At the same time,   applying  the BCFW shift \rf{bcfw1}  to the  momenta  $p_1$ and $p'_1$   in \rf{51} 
and  assuming a similar large $w$ singularity cancellation due to the contact  term contribution   we get  instead 
\begin{equation}
\label{bcfw6}
\hat {\cal A}''(\s,\t,\u)=-  i\frac{ \t}{\s(\s+\t)}F(2\t)  \  .
\end{equation}
The two expressions   (\ref{bcfw55}) and (\ref{bcfw6})  could   agree  provided 
$F(z) = \frac{c}{ z^2}$   but this  contradicts the actual   form of $F$ in  \rf{56}. 

We are thus led to the conclusion that the  cubic 
vertices  \rf{8} with the couplings \rf{29}   which  led to the exchange  amplitude \rf{57} 
 are inconsistent with the condition of 
BCFW    constructibility.


\section{Details of summation  over  spins}

Here we shall  evaluate some sums over spins appearing in  section 6.
The  bubble diagram   contribution to the   scalar self-energy  (\ref{30}) can be rewritten as
\begin{align}
\Sigma (k^2)= 4 g^2 \ll^{-2}
 \int \frac{d^4p}{p^2 (p-k)^2} \Big[2{\cal S}_1 + \frac{(2kp-p^2) (p^2-k^2)}{(2k-p)^2 (k+p)^2}{\cal S}_2\Big] \ ,
\label{appb1}
\end{align}
where 
\be
{\cal S}_1\equiv \sum_{\substack{{m,n=0}\\{m+n=\ev }}}^{\infty}\frac{x^{m}y^{n}}{\left((m+n-1)!\right)^2},\qquad  \qquad
{\cal S}_2 \equiv \sum_{\substack{{m,n=0}\\{m+n=\ev }}}^{\infty} \frac{m^2 n^2\ x^{m}y^{n} }{\left((m+n-1)!\right)^2}\ ,
\label{appb2} \ee 
 and $x\equiv \ha  {\ell^2 (2k-p)^2}, \ \  y \equiv \ha {\ell^2(k+p)^2}$ as in \rf{610}. 
To evaluate ${\cal S}_1$  let us  first introduce the new variables
\begin{equation}\la{b33}
u\equiv xy\ , \qquad\qquad  v = {x\ov y} \ , 
\end{equation}
 and then rearrange the series to sum over $m$ and $n$ obeying $m+n=2r $ with fixed  integer $r$ first
\begin{align*}
{\cal S}_1 = \sum_{\substack{{m,n=0}\\{m+n=2r }}}^{\infty}\frac{u^{{(m+n)}/{2}}v^{{(m-n)}/{2}}}{\left((m+n-1)!\right)^2}=
\sum_{r =0}^{\infty}\frac{u^{r }v^{r }}{\left((2r -1)!\right)^2}\sum_{n=0}^{2r } v^{-n}=
\sum_{r =0}^{\infty}\frac{u^{r }v^{r }}{\left((2r -1)!\right)^2} \frac{1-v^{-(2r +1)}}{1-v^{-1}}.
\end{align*}
After some simple manipulations we find
\begin{align}
\notag
{\cal S}_1&=\frac{v}{v-1}\sum_{r=0}^\infty \frac{x^{2r}}{\left((2r-1)!\right)^2}- 
\frac{1}{v-1}\sum_{r=0}^\infty \frac{y^{2r}}{\left((2r-1)!\right)^2}\\
\label{s1comp}
&=\frac{x^2}{2(x-y)}  \left[I_0(2\sqrt{x})-J_0(2\sqrt{x}) \right]-\frac{y^2}{2(x-y)} \left[ I_0(2\sqrt{y})-J_0(2\sqrt{y})\right]\ ,
\end{align}
where  $J_{0}(z)$ and  $I_{0}(z)$ are the  Bessel functions in  \rf{5.7}. 

Similarly, for  the second sum  we find 
\begin{equation*}
{\cal S}_2 = \sum_{\substack{{m,n=0}\\{m+n=2r}}}^\infty\frac{m^2 n^2 }{\left((m+n-1)!\right)^2}u^{{(m+n)}/{2}}v^{{(m-n)}/{2}}=
\sum_{r =0}^{\infty}\frac{u^{r }v^{r }}{\left((2r -1)!\right)^2}\sum_{n=0}^{2r } n^2 (2r -n)^2v^{-n}.\qquad 
\end{equation*}
Let us  set  $j \equiv 2r=m+n$. Then the  sum over $n$ acquires the following form
\begin{align} &
\sum_{n=0}^j  n^2 (j -n)^2v^{-n} = \frac{v}{(v-1)^5}\Big[ A(v,j ) - v^{-j } B(v,j )\Big] \ , \\
A (v,j )&=j ^2 v^3 +  2j v^3 + v^3 -j ^2v^2+6j v^2 +11v^2 - j ^2v- 6j v+11v+j ^2-2j +1\  , \no \\ 
B(v,j )&=j ^2 v^3 - 2j v^3 + v^3 -j ^2v^2-6j v^2 +11v^2 - j ^2v+6j v+11v+j ^2-2j +1\ .\no 
 \end{align}
 It is  convenient to represent  $A$ and $B$ as 
\begin{align}
 &A(v,j ) = (j -1)(j -2) (v-1)^2 (v+1) + (j -1) (v-1)(v^2-8v-5) +4 (v^2+4v+1),\no \\
&B(v,j ) = (j -1)(j -2) (v-1)^2 (v+1) + (j -1) (v-1)(5v^2+8v-1) +4 v(v^2+4v+1)\ . \no 
\end{align}
Then ${\cal S}_2$ may be written as 
\begin{align}
{\cal S}_2 = \frac{v}{(v-1)^5} \sum_{r =0}^{\infty} \frac{x^{{2r }}}{\left((2r -1)!\right)^2}A(v,2r )-
\frac{v}{(v-1)^5} \sum_{r =0}^{\infty} \frac{y^{{2r }}}{\left((2r -1)!\right)^2}B(v,2r ) \ .
\end{align}
Substituting here  the expressions for $A$ and $B$ we obtain
\begin{align*}
&{\cal S}_2
=\frac{v}{(v-1)^5}  \Big[(v-1)^2 (v+1)\sum_{r =0}^{\infty} \frac{x^{{2r }}}{(2r -1)! (2r -3)!}
-
4v(v^2+4v+1)\sum_{r =0}^{\infty} \frac{y^{{2r }}}{\left((2r -1)!\right)^2} \\
&+
(v-1) (v^2-8v-5)\sum_{r =0}^{\infty} \frac{x^{{2r }}}{(2r -1)! (2r -2)!}
 +4(v^2+4v+1)\sum_{r =0}^{\infty} \frac{x^{{2r }}}{\left((2r -1)!\right)^2} \\
 &-
(v-1)^2 (v+1) \sum_{r =0}^{\infty} \frac{y^{{2r }}}{(2r -1)! (2r -3)!}
-(v-1) (5v^2+8v-1)\sum_{r =0}^{\infty} \frac{x^{{2r }}}{(2r -1)! (2r -2)!}\Big]  .
\end{align*}
Using  \rf{5.7} to perform the summations and expressing  the result  back in terms of $x$ and $y$  in   \rf{b33} 
we find
\begin{align}
\notag
{\cal S}_2 =& \frac{x^3 y (x+y)}{2(x-y)^3}\big[ I_2 (2\sqrt{x})-J_2 (2\sqrt{x})\big]+
\frac{x^2y(x^2 - 8xy-5y^2)\sqrt{x}}{2(x-y)^4}\big[I_1(2\sqrt{x})+J_1(2\sqrt{x})\big]\\
& \qquad\qquad\quad+ \frac{2x^2 y^2 (x^2 + 4xy + y^2)}{(x-y)^5}\big[I_0 (2\sqrt{x})-J_0 (2\sqrt{x})\big]
 +  ( x \leftrightarrow y )\ .
\end{align}
As a result, we get  the  following expression for  \rf{appb1}   which is equivalent to \rf{338},\rf{6.9} 
\begin{align}
\notag
\Sigma (k^2)= 4 g^2 \ll^{-2}  \int \frac{d^4p}{p^2 (p-k)^2} 
\Big(& \Big[\frac{x^2}{x-y}  \big[I_0(2\sqrt{x})-J_0(2\sqrt{x})\big]+(x \leftrightarrow y)\Big]
\\
\notag
&+ \frac{(2kp-p^2) (p^2-k^2)}{(2k-p)^2 (k+p)^2}\Big[\frac{x^3 y (x+y)}{2(x-y)^3}
\big[ I_2 (2\sqrt{x})-J_2 (2\sqrt{x})\big]\\
\label{appb5}
&+\frac{x^2y(x^2 - 8xy-5y^2)\sqrt{x}}{2(x-y)^4}\big[I_1(2\sqrt{x})+J_1(2\sqrt{x})\big]\\
& 
+ \frac{2x^2 y^2 (x^2 + 4xy + y^2)}{(x-y)^5} \big[I_0 (2\sqrt{x})-J_0 (2\sqrt{x})\big] +  ( x \leftrightarrow y )\Big]\Big) \ . 
\no 
\end{align}
Let us note that (\ref{appb5}) is not directly  applicable for $x=y$ 
because intermediate 
summations over powers of $v$ resulted in a spurious pole at $v=1$.
The $x= y$  limit is actually regular as one can show  by 
expanding the  numerators in powers of  $(x-y)$ and checking that all    factors  of  $(x-y)$ in the denominator  get 
cancelled.
 Equivalently,  one may just  evaluate the above sums  explicitly for $v=1$. This gives
\begin{align}
\label{appb6}
{\cal S}_1|_{x=y}=& \ha  (\sqrt{x})^3 \big[I_1(2\sqrt{x})+J_1(2\sqrt{x})\big]+ x \big[I_0(2\sqrt{x})-J_0(2\sqrt{x})\big] \ ,\\
{\cal S}_2|_{x=y} =&\te \frac{1}{60} (\sqrt{x})^7 \big[I_5(2\sqrt{x})+J_5(2\sqrt{x})\big]
+ \fo  x^3\big[I_4(2\sqrt{x})-J_4(2\sqrt{x})\big]
\notag
\\
\no 
 &\te  +
\frac{13}{12}(\sqrt{x})^5 \big[I_3(2\sqrt{x})+J_3(2\sqrt{x})\big]+{3\ov 2} x^2 \big[I_2(2\sqrt{x})-J_2(2\sqrt{x})\big]
\no \\ 
&+\ha (\sqrt{x})^3 \big[I_1(2\sqrt{x})+J_1(2\sqrt{x})\big]\ .  \label{appb7}
\end{align}
As a result,  in the  $x=y$ limit   (which  corresponds according to  \rf{610}   to  
the  $k^2=0$ value of the self-energy
correction  and thus determines the leading UV  $p\to \infty$  behaviour of the integrand in \rf{appb1}) 
  one finds the expression given in \rf{616}.
 

\bibliography{loopsbib}
\bibliographystyle{utphys}
\end{document}

\


\providecommand{\href}[2]{#2}\begingroup\raggedright\endgroup

\end{document}